\documentclass[12pt]{article}
\usepackage{amsmath}
\usepackage{amssymb}
\usepackage{graphicx}
\usepackage{epsfig}
\usepackage{subfigure,float, afterpage}
\usepackage{color}
\usepackage{cite}
\usepackage{caption}

\usepackage[titletoc]{appendix}
 \usepackage[section]{placeins}

\providecommand{\keywords}[1]{\textbf{\textit{Keywords :}} #1}
\usepackage[paper=a4paper,top=1in,bottom=1in,right=1in,left=1in]{geometry}

\begin{document}

%{\makebox[1.cm]
%        \sf hep$\,$-$\,$ph/0705xxx}

%\vspace*{0.5cm}
\setcounter{footnote}{-1}
 \begin{center}
    {\LARGE{\bf{Checking T and CPT violation with sterile neutrino }}}
    \end{center}
\renewcommand{\thefootnote}{\fnsymbol{footnote}}
%
%\renewcommand{\thefootnote}{\it\alph{footnote}}
%
%\vspace*{1cm}
%
{\begin{center}
{{\bf                
                  ${Yogita \enspace Pant}^{1}$, ${Sujata \enspace Diwakar}^{2}$, ${Jyotsna \enspace Singh}^{3}$, ${R.B \enspace Singh}^{4}$ 
                 
               $^{1234}$Department of Physics, University of Lucknow, Lucknow 226007 India
               
                }}
               \end{center}} 
                \vskip 1.2cm
%{\small
              %  \begin{center}
               % $^a$ Harish-Chandra Research Institute, Chhatnag Road, \\	
                %     Jhunsi, Allahabad 211 019, India\\[4mm]
                 %     \end{center}}
\begin{center}
              {\bf{Abstract}}
              \end{center}
Post LSND results, sterile neutrinos have drawn attention and motivated the high energy physics, astronomy and cosmology to probe physics beyond the standard model considering minimal 3+1 (3 active and 1 sterile) to 3+N neutrino schemes. The analytical equations for neutrino conversion probabilities are developed in this work for 3+1 neutrino scheme. Here, we have tried to explore the possible signals of T and CPT violations with four flavor neutrino scheme at neutrino factory. Values of sterile parameters considered in this analysis are taken from two different types of neutrino experiments viz. long baseline experiments and reactor+atmospheric experiments. In this work golden and discovery channels are selected for the investigation of T violation. While observing T violation we stipulate that neutrino factory working at 50 GeV energy have the potential to observe the T violation signatures for the considered range of baselines(3000 km-7500 km). The ability of neutrino factory for constraining CPT violation is enhanced with increase in energy for normal neutrino mass hierarchy(NH). Neutrino factory with the exposure time of 500 kt-yr will be able to capture CPT violation with
 $ \delta c_{31}\geq 3.6\times10^{-23} $ GeV at 3$ \sigma $ level for NH and for IH with $ \delta c_{31}\geq 4\times10^{-23} $ GeV at 3$ \sigma $ level.

%These observables promote setup with baselines beyond 4000 km and energy range 9-12 GeV to check CPT violation. We find that to observe CPT violation at 90\% C.L., the product of sterile angles $ \theta_{24} $ and $ \theta_{34} $ can be constrained  as $\theta_{24}.\theta_{34}\gtrsim 0.006$ $rad^{2}$ with 
%liquid argon detector and with magnetized iron detector the constraint imposed is $\theta_{24}.\theta_{34}\gtrsim 0.007$ $rad^{2}$.\\\vspace{-1ex}
\newpage  

 \section{Introduction}
The standard model of particle physics considers neutrinos to be massless. Sudbury Neutrino Observatory\cite{Ahmad}\cite{Ahmed} gave evidence of neutrino oscillations which was further confirmed by KamLAND experiment \cite{Eguchi}. This landmark research assigned mass to the neutrinos and gave a clear indication of new physics beyond the standard model. A simple stretch in the standard model was able to stand up with the mass of neutrino. 
In neutrino physics the standard three flavour neutrino oscillations can be explained with the help of six parameters namely $\theta_{12}$, $\theta_{13}$, $\theta_{23}$, $\Delta m^{2}_{12}$, $\Delta m^{2}_{31}$  and $\delta_{CP} $. Amongst these six parameters, solar  parameters($ \theta_{12}$, $\Delta m^{2}_{12}$) and atmospheric parameters ($  \theta_{23}, \Delta m^{2}_{31} $)  have been measured with high precision. Furthermore, Daya Bay and RENO reactor experiments have strongly constrained the value of mixing angle $ \theta_{13} $. Now we are in need of such neutrino experiments which can impose tight constraints on the value of $\delta_{CP}$ and mass hierarchy. Some anomalies popped up while observing appearance channel and disappearance channel of $\nu_{e}$ at LSND experiment. While observing  $ \bar{\nu}_{\mu}  \rightarrow \bar{\nu_{e}} $ appearance channel, LSND \cite{C.Athanassopoulos} \cite{Phys1} \cite{Phys2} \cite{Phys3} \cite{Phys4} \cite{Phys6} was the first experiment to publish evidence of a signal at $ \Delta m^{2} \sim 1 eV^{2} $. Later in 2002, MiniBooNE \cite{phy7} \cite{phy8} checked the LSND result for $ \nu_{e} \rightarrow \nu_{\mu}$  ($\bar{\nu}_{e} \rightarrow \bar\nu_{\mu} $) appearance channel. In MiniBooNE experiment, while observing the CCQE events rate through $ \nu_{e}n\rightarrow e^{-}p(\bar{\nu}_{e}p\rightarrow e^{+}n)$ above 475 MeV energy, no excess  events were found but for energies $\textless$ 475 MeV  $ \nu_{e} (\bar{\nu}_{e})$  excess events were observed. In this way, MiniBooNE supported the LSND result. The LEP data\cite{yao}\cite{Nakeamura} advocates the number of weakly interacting light neutrinos, that couple with the Z bosons through electroweak interactions, to be 2.984 $\pm$ 0.008; thus closing the door for more than three active neutrinos. Hence, the heavy neutrino announced by LSND group should be different from these three active neutrinos. This higher mass splitting in the standard three active neutrino model  was accommodated by introducing  sterile neutrinos. Sterile neutrinos carry a new flavor which can mix up with the other three flavors of standard model but they do not couple with W and Z bosons. The number of sterile neutrinos can vary from minimum one to any integer N.\\
Some cosmological evidences like CMB anisotropies \cite{phy14} \cite{phy15} \cite{phy16} \cite{phy17} \cite{phy18} and Big Bang nucleosynthesis \cite{phy19} \cite{phy20} also stood up with the LSND data. The results reported by the combined analysis \cite{phy26} of Baryonic Acoustic Oscillations(BAO) `$H_{0}$+PlaSZ+Shear+RSD' indicated the presence of sterile neutrinos by stipulating the number of effective neutrinos $ N_{eff}$ $\equiv 3.62^{+0.26}_{-0.42} $, $ m_{\nu}^{eff}(sterile) =4.48_{-0.14}^{+0.11} eV $ and  giving preference for $ \bigtriangleup N_{eff} \equiv  N_{eff}- 3.62^{+0.26}_{-0.42} $  at 1.4 $ \sigma $ level and non zero mass of sterile neutrino at 3.4 $ \sigma $ level.
%and $ m_{\nu}^{eff}=4.48_{-0.14}^{+0.11} eV $ for $N_{eff}\equiv N_{eff}-3.046>0$ at 1.4 $ \sigma $.\\
The gallium solar neutrino experiments(gallium anomaly) GALLEX \cite{phy27}, SAGE \cite{phy28} and the antineutrino reactor experiments ($\bar{\nu}_{e} $) like Bugey-3, Bugey-4, Gosgen, Kransnogark, IIL \cite{new1} (reactor anomaly) indicated that electron neutrinos and antineutrinos may disappear at short baselines. Such disappearance can be explained by the presence of at least one massive neutrino (of the order of 1 eV). Thus, these experiments also indicated the presence of sterile neutrino and supported the LSND results. Some constraints imposed by the combined fit of reactor, gallium, solar and $ \nu_{e}C $  scattering data are  $ \bigtriangleup m^{2}_{41}\gtrsim 1 eV^{2}$  and  $0.07 \leq sin^{2}2\nu_{ee} \lesssim 0.09$ at 95\% CL \cite{new2}. Few atmospheric neutrino experiments such as IceCube \cite{phy9}, MINOS \cite{phy10} \cite{phy11} \cite{phy12}, CCFR \cite{phy13} have also imposed strong constraints on sterile  parameters.\\
The four flavors of neutrino can be studied in  either of the two different neutrino mass schemes,  3+1 or 2+2 schemes\cite{caldwell}. For our work we have selected  (3+1) four flavor neutrino mass scheme. In this framework, Maki-Nakagawa-Sakata (MNS)  mixing matrix ($4\times 4$), includes  six mixing angles $\theta_{ij}$, three dirac phases and three majorana  phases. In our analysis majorana phases are not taken into consideration. \\
Neutrino factory \cite{Geer}\cite{Rujula} provides excellent sensitivity to the standard neutrino oscillation parameters and therefore seems to be one of the promising option to explore and reanalyze the global fits for sterile neutrino parameters too. To  mention, it provides a platform to constrain one of the most searched  CP violation in leptonic sector \cite{Albright}\cite{Blonded}. Hence neutrino factory seems to provide a promising environment for the study of T and CPT violation. The neutrino factory set up considered here is based on the International Design Study of Neutrino Factory [IDS-NF] \cite{Donini}\cite{Gavela}. From the measure of $ \Delta P_{CP} $ we can not directly constrain CP phase because the value of $ \Delta P_{CP} $ in the presence of matter will contain in itself some CP odd effects even in the absence of CP phase. Therefore, instead of checking $ \Delta P_{CP}$, variation in $ \Delta P_{T} $ can be studied to probe extent of true CP violation.  We have observed T violation through $\nu_{\mu}\rightarrow \nu_{e}$ golden channel and $\nu_{\mu}\rightarrow \nu_{\tau}$ discovery channel and CPT violation along $\nu_{\mu}\rightarrow \nu_{\mu}$  disappearance channel.\\
 Our work is organized as follows.
 In section 2, we illustrate 3+1 neutrino matrix parametrization. In next section, T violating effects are checked for different channels. In section 4, bounds on CPT violating terms are checked in presence of sterile neutrino. In the last section, we have summarized our study and discussed the results observed.            
\section{Standard Parametrization in 3 + 1 neutrino scheme}
The 3 (active) + 1 (sterile) neutrino scheme can be looked upon as 3+1 or 2+2 scheme depending on the selection of mass ordering of the neutrinos. To check T and CPT violation we have selected 
3+1 scheme for our analysis. In this scheme the flavor eigenstates $ \nu_{\alpha} (\alpha= e,\mu ,\tau, s) $ and mass eigenstates $ \nu_{j} (j=1,2,3,4) $ are related by the given unitary transformation equation \\

\begin{equation}
\begin{pmatrix}
 \nu_{e} \\
 \nu_{\mu} \\
 \nu_{\tau} \\
 \nu_{s} 
\end{pmatrix}= U \begin{pmatrix}
 \nu_{1} \\
 \nu_{2} \\
 \nu_{3} \\
 \nu_{4} 
\end{pmatrix}               
\end{equation}\\

\footnote{A N$\times$ N unitary matrix contains N(N-1)/2 mixing angles and (N-1)(N-2)/2 Dirac type CP violating phases. It will also contain (N-1) number of additional Majorana Phases if the neutrinos are considered as Majorana particles.}\\
Here unitary matrix (U) can be parametrized in terms of six mixing angles($ \theta_{12}, \theta_{13},\\ \theta_{23}, \theta_{14}, \theta_{24}, \theta_{34} $), three Dirac phases $ \delta_{l} $ ($ \delta_{1},\delta_{2},\delta_{3} $) and three majorana phases.
Majorana phases are neglected in our study as they do not affect the neutrino oscillations in any realistically observable way. In principle, there are different parametrization schemes for the neutrino mixing matrix as their order of sub-rotation is arbitrary. Our selection for parametrization of neutrino mixing matrix is

\begin{equation}
U = U_{34}(\theta_{34},0)U_{24}(\theta_{24},0)U_{14}(\theta_{14},0)U_{23}(\theta_{23},\delta_{3})U_{13}(\theta_{13},\delta_{2})U_{12}(\theta_{12},\delta_{1}) 
\end{equation} 
     
where $U_{ij}(\theta_{ij},\delta_{l})$ are the complex rotation matrices in the ij plane, defined as
\begin{equation}
[U_{ij}({\theta}_{ij},{\delta}_{l})]_{pq}=\begin{cases}
cos{\theta}_{ij} &  p = q = i,j\\
1  &  p = q \neq i,j\\
sin{\theta}_{ij} e^{-i\delta_{l}}  &   $p = i , q = j$\\
-sin{\theta}_{ij} e^{i\delta_{l}}  &   $p = j , q = i$\\
0  &     \text{otherwise}
\end{cases}
\end{equation}
The order of rotation  between 14 and 23  is arbitrary since these matrices commute. When neutrinos pass through the earth matter, the charge current interactions (CC) of $\nu_{e}$ and neutral current interactions (NC) of $\nu_{e}, \nu_{\mu}, \nu_{\tau}$  with the matter give rise to a CC and NC potentials $V_{e}$ and $V_{n}$ respectively. While studying the sterile neutrinos, potential $ V_{n} $ can not be neglected. The effective CPT violating hamiltonian ($ H_{f} $) of neutrinos can be expressed as  
\begin{center}
\begin{equation}
\begin{split}
 H_{f} & =\dfrac{1}{2E} [U\begin{pmatrix}
0 & 0 & 0 & 0\\
0 & \Delta m_{21}^{2} & 0 & 0\\
0 & 0 & \Delta m_{31}^{2} & 0\\
0 & 0 & 0 & \Delta m_{41}^{2} \end{pmatrix} U^{\dagger}+ U_{b}\begin{pmatrix}
0 & 0 & 0 & 0\\
0 & \delta{c_{21}}.2E & 0 & 0\\
0 & 0 & \delta{c_{31}}.2E & 0\\
0 & 0 & 0 & \delta{c_{41}}.2E \end{pmatrix} U_{b}^{\dagger}\\
&\quad +\begin{pmatrix}
A_{e}+A_{n} & 0 & 0 & 0\\
0 & A_{n} & 0 & 0\\
0 & 0 & A_{n} & 0\\
0 & 0 & 0 & 0 \end{pmatrix}] 
\end{split}
\end{equation}
\end{center} 
Here $A_{e(n)}=2EV_{e(n)}$,  $V_{e}=\sqrt{2}G_{F}N_{e}$ and $V_{n}=-G_{F}N_{n}/\sqrt{2}$. $G_{F}$ is the Fermi constant, $N_{e}$ and $N_{n}$ are the number density of electrons and neutrons respectively with $N_{e}\simeq N_{n}$ in earth matter. The $ \delta c_{ij}$'s  are CPT violating terms. Different angular values for unitary matrix $U_{b}$ are checked in \cite{animesh}. In our work we have considered $U = U_{b}$. Hamiltonian $H_{f}$ can be diagonalized  to $ H_{D} $ by an unitary matrix $\tilde U$. This can be expressed as   \vspace{-5ex}
\begin{center}
\begin{equation}
 H_{D} = \tilde{U}^{\dagger} H_{f}\tilde{U} 
\end{equation}
\end{center}
The matrix  elements  $[ H_{D}]_{ii}$  will represent  the eigenvalues of $H_{f}$. Full analytical expressions for neutrino oscillation probabilities are developed in this work by using time independent perturbation theory. In an attempt to apply perturbation we have defined few oscillation parameters in terms of perturbative parameter $\eta$, where $ \eta = 0.18 $ . The neutrino oscillation parameters can be rewritten as  \\
$\theta_{14} \equiv \chi_{14} \eta$ \\
$\theta_{24} \equiv \chi_{24} \eta$\\
$\theta_{34} \equiv \chi_{34} \eta$ \\
$\theta_{13} \equiv \chi_{13} \eta$ \\
$\hat{\theta_{23}}\simeq \theta_{23} - 1/\sqrt{2} \equiv \chi_{23} \eta$\\\\
We treat $\dfrac{\Delta m^{2}_{21}}{\Delta m^{2}_{31}} \approx O(\eta^{2})$. Now the Hamiltonian $H_{f}$ can be written as  \vspace{-5ex}
\begin{center}
\begin{equation}
H_{f} = \frac{\Delta m^{2}_{31}}{2E} \left[ H_{0} + H_{1}\eta + H_{2}\eta^{2} + O(\eta^{3})\right] 
\end{equation}
\end{center} 
where $H_{0}$, $H_{1}$ and $H_{2}$ are the hamiltonians corresponding to zeroth, first and second order in $\eta$ respectively. 
The evolution equation for neutrino oscillation probability is defined as 
\vspace{-5ex}

\begin{equation}
P_{\alpha\beta} = \mid S_{\beta\alpha}(t,t_{0})\mid^{2} 
\end{equation}
 \vspace{-0.9ex}
where $S(t,t_{0})$ is the evolution matrix of neutrino which is also called oscillation probability amplitude\vspace{-8ex}
\begin{center}
\begin{equation}
\mid\nu(t)> = S(t,t_{0})\mid\nu(t_{0})>
\end{equation}
\end{center}
The evolution matrix of neutrinos in terms of eigenvalues of $H_{D}$ can be written as 
\begin{equation}
S_{\beta \alpha}(t,t_{0}) =  \displaystyle\sum_{i=1}^{4}(\tilde{U}_{\alpha i})^{\ast} \tilde{U}_{\beta i}e^{-iE_{i}L} 
\end{equation}
where $L\equiv t-t_{0} $\\
From equation (7) the neutrino oscillation probability  $ P_{\alpha \beta} $ from flavor $\alpha$ to flavor $\beta$ can be written as \vspace{-5ex} 

\begin{equation}
P_{\alpha \beta} = \mid \displaystyle\sum_{i=1}^{4} (\tilde{U}_{\alpha i})^{*}(\tilde{U}_{\beta i})e^{-iE_{i}L}\mid^{2} 
\end{equation}

This is the general form of equation for neutrino oscillation probability.

\section{T violation in (3+1) framework}
In neutrino oscillations the flavor conversion probabilities from flavor $\alpha$ to flavor $\beta$  can be written as \vspace{-4ex}
\begin{center}
\begin{equation}
{P}_{\nu_{\alpha}\longrightarrow \nu_{\beta}} = {\delta}_{ij} - 4 \displaystyle\sum_{i>j} Re\left[ {\tilde{U}}_{\alpha i} {\tilde{U}}_{\alpha j}^{*} {\tilde{U}}_{\beta i}^{*} {\tilde{U}}_{\beta j}\right] {\sin}^{2}{{\Delta}_{ij}} + 2 \displaystyle\sum_{i>j} Im\left[{\tilde{U}}_{\alpha i} {\tilde{U}}_{\alpha j}^{*} {\tilde{U}}_{\beta i}^{*} {\tilde{U}}_{\beta j}\right] \sin 2{\Delta}_{ij} 
\end{equation}
\end{center}
Redefining the above probability equation as sum of $P_{CP-even}$ and $P_{CP-odd}$ terms \vspace{-6ex}
\begin{center}
\begin{equation}
P_{\alpha\beta} = P_{(\nu_{\alpha}\rightarrow \nu_{\beta})} = {P}_{CP-even}(\nu_{\alpha}\rightarrow \nu_{\beta}) + {P}_{CP-odd}(\nu_{\alpha}\rightarrow \nu_{\beta})
\end{equation}
\end{center}
CP even terms are CP conserving and  can be written as\vspace{-6ex}
\begin{center}
\begin{equation}
 P_{(\nu_{\alpha}\rightarrow \nu_{\beta})} = P_{(\bar{\nu}_{\alpha}\rightarrow\bar{\nu}_{\beta})}
                                          = \delta_{ij} - 4\displaystyle\sum_{i>j}Re (\tilde{U}_{\alpha i}\tilde{U}_{\beta i}^{\ast}\tilde{U}_{\alpha j}^{\ast}\tilde{U}_{\beta j} )\sin^{2}{\Delta_{ij}} 
\end{equation}
\end{center}
CP odd term are CP violating and  can be written as \vspace{-6ex}
\begin{center}
\begin{equation}
 P_{(\nu_{\alpha}\rightarrow\nu_{\beta})} = -P_{(\bar{\nu}_{\alpha}\rightarrow \bar{\nu}_{\beta})}
                                          = 2\displaystyle\sum_{i>j}Im (\tilde{U}_{\alpha i}\tilde{U}_{\beta i}^{\ast}\tilde{U}_{\alpha j}^{\ast}\tilde{U}_{\beta j} )\sin{2\Delta_{ij}} 
\end{equation}
\end{center}
Assuming CPT to be conserved, the magnitude of CP violation $(\Delta P_{CP})$ will be equal to the magnitude of T violation $(\Delta P_{T})$, i.e. \vspace{-6ex}
\begin{center}
\begin{equation}
\mid \Delta P_{CP}\mid =  \mid \Delta P_{T}\mid
\end{equation}
\end{center}
therefore we can write
\vspace{-8ex}
\begin{center}
\begin{equation}
 P_{(\nu_{\alpha}\rightarrow \nu_{\beta})}-P_{(\bar{\nu_{\alpha}}\rightarrow \bar{\nu_{\beta}})}\equiv P_{(\nu_{\alpha}\rightarrow \nu_{\beta})}-P_{(\nu_{\beta}\rightarrow \nu_{\alpha})}
\end{equation}
\end{center}
 When neutrinos passes through the earth matter, the interaction of neutrinos with matter gives rise to an extra potential. This potential is positive for neutrinos and negative for antineutrinos leading to different eigenvalues of hamiltonian for them. Further, this difference in hamiltonian for $ \nu $'s and $ \bar{\nu} $'s give rise to fake(extrinsic) CP violation. 
 Hence, check on T violation appears to be a better choice in the presence of matter. From equation (16)  the T violation can be looked upon as  \vspace{-7ex}
\begin{center}
\begin{equation}
(\Delta P_{T})_{\alpha\beta} = P(\nu_{\alpha}\rightarrow \nu_{\beta})-P(\bar{\nu_{\alpha}}\rightarrow \bar{\nu_{\beta}})\equiv 4\displaystyle\sum_{i>j}Im (\tilde{U}_{\alpha i}\tilde{U}_{\beta i}^{\ast}\tilde{U}_{\alpha j}^{\ast}\tilde{U}_{\beta j} )\sin{2\Delta_{ij}}
\end{equation}
\end{center}
If we consider $\tilde{U}_{\alpha j}\tilde{U}_{\beta j}^{\ast} = \tilde{V}_{j}^{\alpha \beta} $ and $\Delta_{ij} = 2 \Delta\tilde{E}_{jk}/L = 2 (\tilde{E_{j}}-\tilde{E_{k}})/L$ the above equation becomes \vspace{-7ex}
\begin{center}
\begin{equation} 
(\Delta P_{T})_{\alpha\beta}\equiv 4\displaystyle\sum_{j<k} Im({\tilde{V_{j}}^{\beta\alpha}}{\tilde{V_{k}}^{\beta\alpha^{\ast}}})\sin{(\Delta\tilde{E}_{jk}L)}
\end{equation}
\end{center}
 The term $Im({\tilde{V_{j}}^{\beta\alpha}}{\tilde{V_{k}}^{\beta\alpha^{\ast}}})$ is known as Jarlskog factor and $\tilde{E}_{j}'s$ are energy eigenvalues of hamiltonian in matter.
\begin{equation}
\begin{split}
(\Delta P_{T})_{\alpha\beta} & =4 Im({\tilde{V_{1}}^{\beta\alpha}}{\tilde{V_{2}}^{\beta\alpha^{\ast}}})\sin{(\Delta\tilde{E}_{12}L)} + 4 Im({\tilde{V_{1}}^{\beta\alpha}}{\tilde{V_{3}}^{\beta\alpha^{\ast}}})\sin{(\Delta\tilde{E}_{13}L)}\\
&\quad + 4 Im({\tilde{V_{2}}^{\beta\alpha}}{\tilde{V_{3}}^{\beta\alpha^{\ast}}})\sin{(\Delta\tilde{E}_{23}L)}
+ 4 Im({\tilde{V_{1}}^{\beta\alpha}}{\tilde{V_{4}}^{\beta\alpha^{\ast}}})\sin{(\Delta\tilde{E}_{14}L)}\\
&\quad + 4 Im({\tilde{V_{2}}^{\beta\alpha}}{\tilde{V_{4}}^{\beta\alpha^{\ast}}})\sin{(\Delta\tilde{E}_{24}L)}
 + 4 Im({\tilde{V_{3}}^{\beta\alpha}}{\tilde{V_{4}}^{\beta\alpha^{\ast}}})\sin{(\Delta\tilde{E}_{34}L)} 
\end{split} 
\end{equation}
The energy eigenvalues in matter can be connected to the energy eigenvalues in vacuum by the relations $\tilde{E}_{1} = \Delta E_{31}$, $\tilde{E}_{2} = 0$, $\tilde{E}_{3} = A_{e}$ and $\tilde{E}_{4} = \Delta E_{41}$. The terms $\tilde{V_{1}}^{\alpha \beta}$,$\tilde{V_{2}}^{\alpha \beta}$, $\tilde{V_{3}}^{\alpha \beta}$, $\tilde{V_{4}}^{\alpha \beta}$ in matter can be calculated with the help of the terms $ V_{1}^{\alpha \beta} $, $V_{2}^{\alpha \beta}$,$V_{3}^{\alpha \beta}$, $V_{4}^{\alpha \beta}$ in vacuum (where ${V}_{j}^{\alpha \beta} = {U}_{\alpha j}{U}_{\beta j}^{\ast} $) using the following expressions \cite{yasuda}.  \vspace{-6ex}
\begin{center}
\begin{equation}
\tilde{V_{1}}^{\beta\alpha} = -\Delta \tilde{E}_{21}^{-1}\Delta \tilde{E}_{21}^{-1}\lbrace V_{4}^{\beta\alpha}\tilde{E}_{2}\tilde{E}_{3} + (\tilde{E}_{2} + \tilde{E}_{3})R^{\beta \alpha} + S^{\beta\alpha}\rbrace
\end{equation}
\end{center}

\begin{center}
\begin{equation}
\tilde{V_{2}}^{\beta\alpha} = \Delta \tilde{E}_{21}^{-1}\Delta \tilde{E}_{32}^{-1}\lbrace V_{4}^{\beta\alpha}\tilde{E}_{3}\tilde{E}_{1} + (\tilde{E}_{3} + \tilde{E}_{1})R^{\beta \alpha} + S^{\beta\alpha}\rbrace
\end{equation}
\end{center}
\vspace{-9ex}
\begin{center}
\begin{equation}
\tilde{V_{3}}^{\beta\alpha} = -\Delta \tilde{E}_{31}^{-1}\Delta \tilde{E}_{32}^{-1}\lbrace V_{4}^{\beta\alpha}\tilde{E}_{1}\tilde{E}_{2} + (\tilde{E}_{1} + \tilde{E}_{2})R^{\beta \alpha} + S^{\beta\alpha}\rbrace
\end{equation}
\end{center}
\vspace{-9ex}
\begin{center}
\begin{equation}
\tilde{V_{4}}^{\beta\alpha} = V_{4}^{\beta\alpha}
\end{equation}
\end{center}  \vspace{-5ex} 
where  \vspace{-5ex} 
\begin{center}
\begin{equation}
R^{\beta\alpha} = \lbrace A(V_{4}^{ee} + V_{4}^{SA}/2)-A_{\alpha\alpha}-A_{\beta\beta}\rbrace V_{4}^{\beta\alpha} + \Delta E_{31}V_{3}^{\beta\alpha} + \Delta E_{21}V_{2}^{\beta\alpha}
\end{equation}
\end{center}
\vspace{-8ex}

\begin{equation}
\begin{split}
S^{\beta\alpha} & = V_{4}^{\beta\alpha}\lbrace A_{\alpha\alpha}^{2} + A_{\alpha\alpha}A_{\beta\beta}
 + A_{\beta\beta}^{2}-A(A_{\alpha\alpha}+A_{\beta\beta})(V_{4}^{ee}+V_{4}^{SS}/2)\rbrace\\ &\quad-\Delta E_{31}(\Delta E_{31}+A_{\alpha\alpha}+A_{\beta\beta})V_{3}^{\beta\alpha}
 -\Delta E_{21}(\Delta E_{21}+A_{\alpha\alpha}+A_{\beta\beta})V_{2}^{\beta\alpha}\\
 &\quad+A\Delta E_{31}(V_{4}^{\beta e}V_{3}^{e\alpha}+V_{3}^{\beta e}V_{4}^{e\alpha}
+V_{4}^{\beta s}V_{3}^{s\alpha}+ V_{3}^{\beta s}V_{4}^{s\alpha})\\
&\quad+A \Delta E_{21}(V_{4}^{\beta e}V_{2}^{e\alpha}+V_{2}^{\beta e}V_{4}^{e\alpha}
+V_{4}^{\beta s}V_{2}^{s\alpha}+V_{2}^{\beta s}V_{4}^{s\alpha})
 \end{split} 
\end{equation}
$A_{\alpha} = A_{e}\delta_{\alpha e} - A_{n}\delta_{\alpha s}$ is the diagonal element of the matter potential matrix in four neutrino scheme and A is the diagonal element of matter potential matrix in three neutrino scheme. Since T violating effects can only be studied in appearance channels so $\alpha\neq \beta$. In an effort to put constraints on $ \bigtriangleup P_{T} $  we have studied two appearance channels. These are  $\nu_{e}\rightarrow \nu_{\mu}$ (golden channel) and $\nu_{\mu}\rightarrow \nu_{\tau}$ (discovery channel).\\
The T violation probability difference expression for the golden channel can be expressed as \vspace{-4ex}
\begin{center}
\begin{equation}
\begin{split}
(\Delta P_{T})_{\mu e} & = \frac{-4}{({\Delta m_{31}^{2}/2E})^{2}\frac{A_{e}}{2E}(\frac{A_{e}}{2E}-\Delta m_{31}^{2}/2E)} \times\\
&\quad
[\frac{A_{e}}{2E}R^{e\mu} + S^{e\mu}][s_{14}c_{14}s_{24}\frac{A_{e}}{2E}\Delta m_{31}^{2}/2E + (\frac{A_{e}}{2E} + \Delta m_{31}^{2}/2E)R^{e\mu} + S^{e\mu}]\sin{\Delta m_{31}^{2}L/2E} \\
&\quad
+ \frac{4}{(-\Delta m_{31}^{2}/2E)\frac{A_{e}}{2E}(\frac{A_{e}}{2E}-\Delta m_{31}^{2}/2E)^{2}} \times\\
&\quad
[\frac{A_{e}}{2E}R^{e\mu} + S^{e\mu}][(\Delta m_{31}^{2}/2E)R^{e\mu} + S^{e\mu}]\sin{(\Delta m_{31}^{2}/2E-\frac{A_{e}}{2E})L} + \\
&\quad
\frac{4}{(-\Delta m_{31}^{2}/2E)(\frac{A_{e}}{2E}-\Delta m_{31}^{2}/2E)(\frac{A_{e}}{2E})^{2}}[s_{14}c_{14}s_{24}\frac{A_{e}}{2E}\Delta m_{31}^{2}/2E + (\frac{A_{e}}{2E} + \Delta m_{31}^{2}/2E)R^{e\mu} \\
&\quad+ S^{e\mu}]
\times[(\Delta m_{31}^{2}/2E) R^{e\mu} + S^{e\mu}]\sin{\frac{A_{e}}{2E}L}
\end{split} 
\end{equation}
\end{center} 
Since large value of $\Delta m_{41}^{2}$  gives rise to rapid oscillations, hence $\Delta m_{41}^{2}$ terms can be averaged out. Solving the above expression up to the power $s_{ij}^{4}$ we get \vspace{-4ex}
\begin{center}
\begin{equation}
\begin{split}
(\Delta P_{T})_{\mu e}  & = 4c_{13}c_{14}^{2}c_{24}s_{13}s_{23}s_{14}s_{24}\sin{(\delta_{2}-\delta_{3})}\frac{\Delta_{e}}{(\Delta_{e}-\Delta_{31})}\sin{\Delta m_{31}^{2}L/2E} \\
&\quad+4c_{13}c_{14}^{2}c_{24}s_{13}s_{23}s_{14}s_{24}\sin{(\delta_{2}-\delta_{3})}\frac{\Delta_{31}^{2}}{\Delta_{e}(\Delta_{e}-\Delta_{31})}\sin{{2\Delta_{e}}}
\end{split}
\end{equation}
\end{center}
\begin{center}
\begin{equation}
\begin{split}
(\Delta P_{T})_{\mu e}  & = 4c_{13}c_{14}^{2}c_{24}s_{13}s_{23}s_{14}s_{24}\sin{(\delta_{2}-\delta_{3})}[ \frac{\Delta_{e}}{(\Delta_{e}-\Delta_{31})}\sin{\Delta m_{31}^{2}L/2E} \\
&\quad + \frac{\Delta_{31}^{2}}{\Delta_{e}(\Delta_{e}-\Delta_{31})}\sin{{2\Delta_{ e}}}] 
\end{split}
\end{equation}
\end{center}
$ \Delta_{e} = A_{e}L/4E $ is matter dependent term. The change in $ \Delta_{e} $ will change the value of $( \Delta P_{T})_{\mu e} $. \\
Further we have developed equation of $ \Delta P_{T} $ for discovery channel. The discovery channel is not very useful in the standard three neutrino flavor framework, nevertheless while studying physics beyond three active neutrino flavor framework, it becomes very important. For discovery channel($\nu_{\mu}\rightarrow \nu_{\tau}$) the probability difference is given as \vspace{-6ex}
\begin{center}
\begin{equation}
\begin{split}
(\Delta P_{T})_{\mu \tau} & = \frac{-4}{({\Delta m_{31}^{2}/2E})^{2}\frac{A_{e}}{2E}(\frac{A_{e}}{2E}-\Delta m_{31}^{2}/2E)}\times \\
&\quad
[\frac{A_{e}}{2E}R^{\tau\mu} + S^{\tau\mu}][c_{14}^{2}c_{24}s_{24}s_{34}\frac{A_{e}}{2E}\Delta m_{31}^{2}/2E + (\frac{A_{e}}{2E} + \Delta m_{31}^{2}/2E)R^{\tau\mu} + S^{\tau\mu}]\sin{\Delta m_{31}^{2}L/2E} \\
&\quad +
\frac{4}{(-\Delta m_{31}^{2}/2E)\frac{A_{e}}{2E}(\frac{A_{e}}{2E}-\Delta m_{31}^{2}/2E)^{2}} \times \\
&\quad
[\frac{A_{e}}{2E}R^{\tau\mu} + S^{\tau\mu}][(\Delta m_{31}^{2}/2E)R^{\tau\mu} + S^{\tau\mu}]\sin{(\Delta m_{31}^{2}/2E-\frac{A_{e}}{2E})L} + \\
&\quad
\frac{4}{(-\Delta m_{31}^{2}/2E)(\frac{A_{e}}{2E}-\Delta m_{31}^{2}/2E)(\frac{A_{e}}{2E})^{2}}[c_{14}^{2}c_{24}s_{24}s_{34}\frac{A_{e}}{2E}\Delta m_{31}^{2}/2E + (\frac{A_{e}}{2E} + \Delta m_{31}^{2}/2E)R^{\tau\mu} \\
&\quad + S^{\tau\mu}]
\times[(\Delta m_{31}^{2}/2E) R^{\tau\mu} + S^{\tau\mu}]\sin{\frac{A_{e}}{2E}L}  
\end{split} 
\end{equation}
\end{center}    
Solving the above expression up to the power $s_{ij}^{4}$ we get, \vspace{-6ex}
\begin{center}
\begin{equation}
\begin{split}
(\Delta P_{T})_{\mu \tau} & = 4c_{13}^{2}c_{14}^{2}c_{23}c_{24}^{2}c_{34}s_{23}s_{24}s_{34}\sin{(\delta_{3})}\frac{\Delta_{e}}{\Delta_{31}}\sin{\Delta m_{31}^{2}L/2E} \\
&\quad-4c_{13}^{2}c_{14}^{2}c_{23}c_{24}^{2}c_{34}s_{23}s_{24}s_{34}\sin{(\delta_{3})}\frac{\Delta_{31}}{\Delta_{e}}\sin{{2\Delta_{e}}}
\end{split}
\end{equation}
\end{center}
\begin{center}
\begin{equation}
\begin{split}
(\Delta P_{T})_{\mu \tau} & = 4c_{13}^{2}c_{14}^{2}c_{23}c_{24}^{2}c_{34}s_{23}s_{24}s_{34}\sin{(\delta_{3})}[ \frac{\Delta_{e}}{\Delta_{31}}\sin{\Delta m_{31}^{2}L/2E} \\
&\quad-\frac{\Delta_{31}}{\Delta_{e}}\sin{2 \Delta_{e}}] 
\end{split}
\end{equation}
\end{center}
%The term $[{\Delta_{e}}/{\Delta_{31}}\sin{\Delta m_{31}^{2}L/2E}-{\Delta_{31}}/{\Delta_{e}}\sin{A_{e}L}]$ in equation (32) is of O(1), so neglecting this term $ \Delta P_{T} $ for discovery channel becomes \vspace{-4ex}
%\begin{center}
%\begin{equation}
%(\Delta P_{T})_{\mu\tau} = \Delta P_{\mu \tau} = 4c_{13}^{2}c_{14}^{2}c_{23}c_{24}^{2}c_{34}s_{23}s_{24}s_{34}\sin{(\delta_{3})}
%\end{equation}
%\end{center}
The $ \Delta P_{T} $ for three neutrino framework \cite{fermilab}  is given by \vspace{-4ex}
\begin{center}
\begin{equation}
\Delta (P_{T})_{3\times 3} \approx 4 c_{12}c_{13}^{2}c_{23}s_{12}s_{13}s_{23}\sin{\delta}
\end{equation}
\end{center}
Keeping the best fit values of neutrino oscillation parameters and assigning maximum value to dirac phases i.e keeping  mod of sin of dirac phases to be unity will lead us to maximum value of $ \Delta P_{T} $. This assumption will render the maximum limit on the bounds which can be imposed on T violation
 arising due to the presence of dirac phases if all other oscillation parameters are known with utmost accuracy. From equation (32) gives the value of $ (\Delta P_{T})_{max}=0.137 $  for three neutrino flavor framework \cite{ref1}.  This value is independent of the selection of probing channel and presence of matter effects. Whereas in 4 flavor framework it will depend on the selection of channel through which we want to probe CP or T violation and it will vary with matter effects too. Within 4 flavor neutrino framework the magnitude of $ \Delta P_{T} $ will depend on active flavor neutrino mixing angles (known with accuracy), sterile neutrino mixing angles (still needs better bounds), matter effects, baseline, energy and dirac phases (not known) . Imposition of constraints on dirac phase (three neutrino flavor) or phases (four neutrino flavor) is still in research phase. 
 
\FloatBarrier
\begin{figure}[htbp]
\begin{minipage}[b]{0.55\linewidth}
\centering
\includegraphics[width=\textwidth]{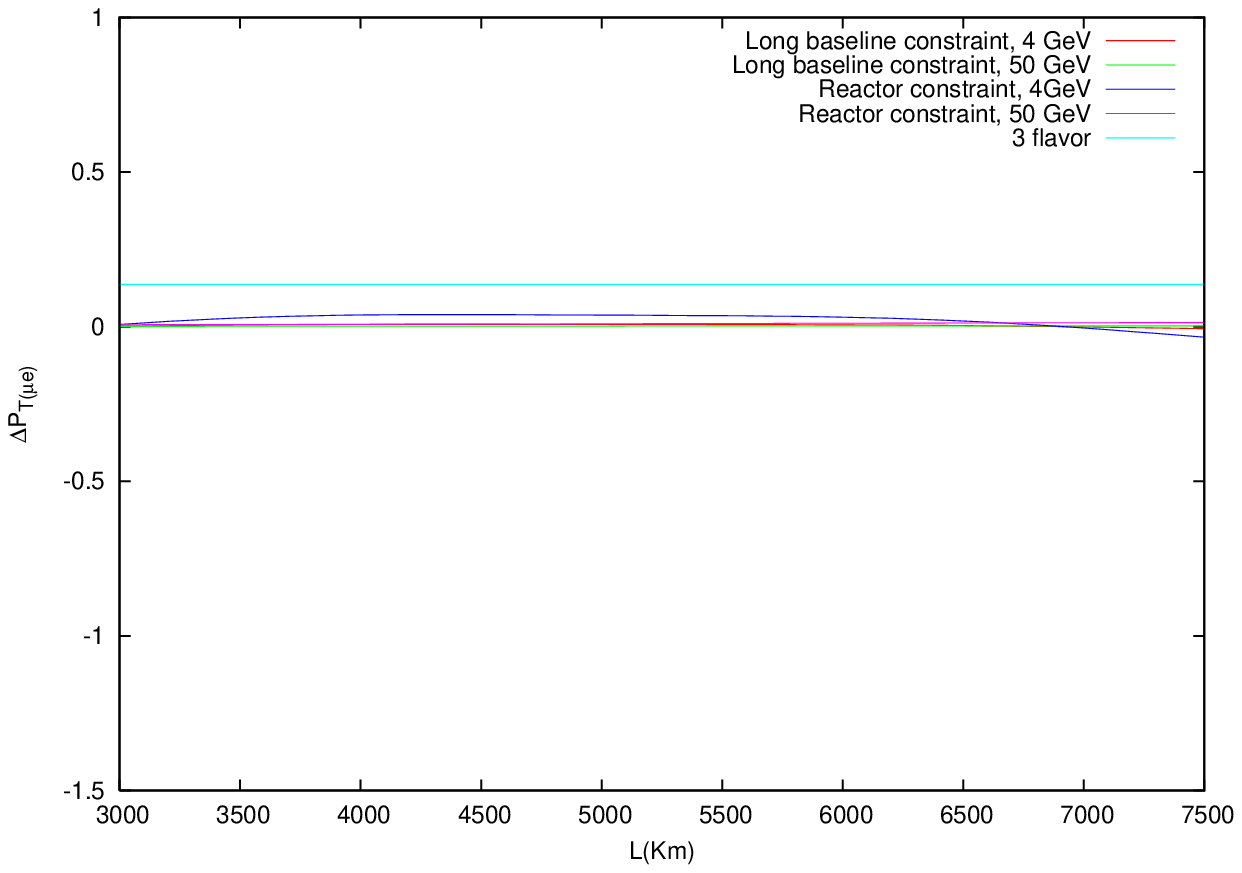}
\subfigure{(a)Variation in $ (\Delta P_{T})_{\mu e} $ along the baseline}
\end{minipage}
\hspace{0.0cm}
\begin{minipage}[b]{0.55\linewidth}
\centering
\includegraphics[width=\textwidth]{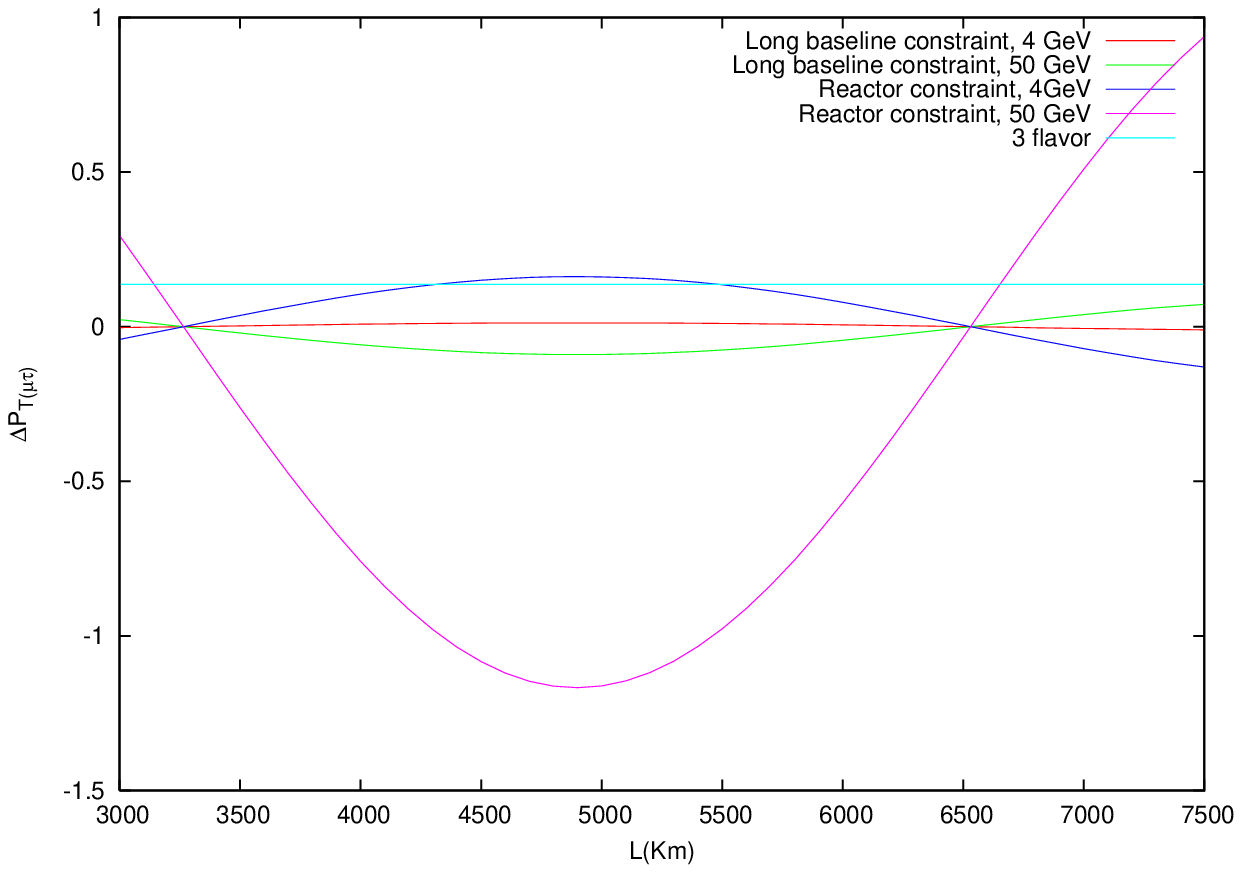}
\subfigure{(b)Variation in $(\Delta P_{T})_{\mu \tau}$ along the baseline}
\end{minipage}
\caption{The variation in $ \Delta P_{T} $ with baseline for two different  appearance channels  }
\end{figure}
The probability differences  $ (\Delta P_{T})_{4\times 4} $ in neutrino sector are represented by equations (28) and (31) for $ \nu_{\mu} \rightarrow \nu_{e} $ and $ \nu_{\mu} \rightarrow \nu_{\tau}  $ channels respectively. The values of sterile parameters used in the above mentioned equations are taken from\\
(i) long baseline neutrino oscillation experiments \cite{meloni} \\$\theta_{14}\lesssim 6.7^{\circ}$, $\theta_{24}\lesssim 3.3^{\circ}$ and $\theta_{34}\lesssim 6.3^{\circ}$    \\
(ii) short baseline reactor and atmospheric experiments \cite{arman} \cite{migliozzi} \\ $\theta_{14}\lesssim 10^{\circ}$, $\theta_{24}\lesssim 12^{\circ}$ and $\theta_{34}\lesssim 28^{\circ}$  \\
The variations in $ (\Delta P_{T})_{\mu e} $ and $ (\Delta P_{T})_{\mu \tau} $ are checked along the baseline for two different energies i.e 4 GeV and 50 GeV. Plot (a) of Figure 1 reflects a very small variation in $ (\Delta P_{T})_{4\times4} $ in comparison to $ (\Delta P_{T})_{3\times3} $ for golden channel when it is checked with two different energies and two sets of sterile parameter values. Plot (b) of Figure 1 reflects a reasonable variation in $ (\Delta P_{T})_{4\times4} $ in comparison to $ (\Delta P_{T})_{3\times3} $ for discovery channel when it is checked for 50 GeV energy and sterile parameters are taken from reactor+atmospheric experiments. The $ \nu_{\mu} \longrightarrow \nu_{\tau} $ comes up as the promising channel to observe the signatures of T violation. From the analysis we conclude that neutrino factory operating at 50 GeV has the potential to capture the signatures of T violation through  $\nu_{\mu} \longrightarrow \nu_{\tau}  $ channel if true sterile parameter values are equal to that taken from reactor+atmospheric experiments. At the same time if the upcoming neutrino experimental setups captures $ \Delta P_{T} $ value  above 0.137 then  we can stipulate presence of some new physics beyond three active flavor neutrino physics which is responsible for the enlargement observed in  $ \Delta P_{T} $ value.
\section{CPT violation in 3 + 1 scheme}
CPT invariance is one of the most fundamental symmetries of nature. CPT conservation indicates the invariance in the properties of physical quantities under the discrete transformations such as charge conjugation (C), parity inversion (P) and time reversal (T) along with the invariance under lorentz transformation. CPT invariance is one of the symmetries of local quantum field theory which implies that there is an important relation between CPT invariance and Lorentz invariance. If CPT invariance is violated, Lorentz invariance must violate but if Lorentz invariance is violated it is not necessary that CPT invariance must violate. In our work $ {\nu_{\mu}\rightarrow \nu_{\mu}} $  disappearance channel is probed to check  CPT violation. The CPT violating probability difference can be written as \vspace{-8ex}
\begin{center}
\begin{equation}
 \bigtriangleup P_{\alpha \beta}^{CPT}=P_{\alpha \beta} -P_{\bar{\beta} \bar{\alpha}} 
\end{equation}
\end{center}
The intrinsic CPT violation arises due to the violation of CPT invariance theorem. A hamiltonian $ H_{f} $ containing CPT violating terms is defined by equation (4) and the general form of neutrino oscillation probability is mentioned in equation (10).
The terms $\tilde{U}_{\alpha i}, {\tilde{U}}_{\alpha j}^{*}, {\tilde{U}}_{\beta i}^{*}$  and ${\tilde{U}}_{\beta j}$ of the expression (10) are the elements of the unitary matrix $\tilde{U}$. The construction of unitary matrix $\tilde{U}$ with the help of eigenvalues and eigenvectors of hamiltonian $ H_{0} $,$ H_{1}$ and $ H_{2} $ is mentioned in the Appendix. After the formation of unitary matrix $ \tilde{U} $, we have developed the neutrino oscillation probability equations up to second order in $\eta$. Since $\Delta m_{41}^{2}$ is large, so we average out the effects produced due to $\Delta m_{41}^{2}$ in the probability equations. Neutrino oscillation probabilities for different oscillation channels containing  CPT violating parameters can be developed  as \vspace{-4ex}
\begin{center}
\begin{equation}
P_{ee} =1-2{\theta_{14}}^{2}-4{\theta_{13}}^{2}(\Delta_{31} + {\delta c}_{31}L/2)^{2}\dfrac{\sin^{2}{(\Delta_{31}+{\delta c}_{31}L/2-\Delta_{e})}}{(\Delta_{31}+ {\delta c}_{31}L/2-\Delta_{e})^{2}}
\end{equation}
\end{center}
\vspace{-4ex}
\begin{center}
\begin{equation}
P_{e\mu}=P_{e\tau}=2 \theta_{13}^{2}(\Delta_{31} + {\delta c}_{31}L/2)^{2}\dfrac{\sin^{2}{(\Delta_{31}+ {\delta c}_{31}L/2-\Delta_{e})}}{(\Delta_{31}+ {\delta c}_{31}L/2-\Delta_{e})^{2}}
\end{equation}
\end{center}
\vspace{-4ex}
\begin{center}
\begin{equation}
\begin{split}
P_{\mu \mu} & = 1-2\theta^{2}_{24}\cos^{2}{(\Delta_{31}+ {\delta c}_{31}L/2)}-(1-8\hat{\theta_{23}^{2}})sin^{2}{(\Delta_{31}+ {\delta c}_{31}L/2)}\\
&\quad+(c_{12}^{2}\Delta_{12}-2\theta_{24}\theta_{34}\cos{\delta_{3}}\Delta_{n})\sin{2(\Delta_{31}+ {\delta c}_{31}L/2)}\\
&\quad +\dfrac{\theta_{13}^{2} (\Delta_{31}+ {\delta c}_{31}L/2)}{(\Delta_{31}+ {\delta c}_{31}L/2-\Delta_{e})^{2}}[2(\Delta_{31}+\delta c_{31}L/2)\sin{\Delta_{e}}\cos{(\Delta_{31}+\delta c_{31}L/2)}\\
&\quad
\sin{(\Delta_{31}+\delta c_{31}L/2-\Delta_{e})}-(\Delta_{31}+\delta c_{31}L/2-\Delta_{e})\Delta_{e}\sin{2\Delta_{31}}+\delta c_{31}L/2]
\end{split}
\end{equation}
\end{center}
\vspace{-4ex}
\begin{center}
\begin{equation}
\begin{split}
P_{\mu\tau} & =\sin^{2}{(\Delta_{31}+ {\delta c}_{31}L/2)}-(8\hat{\theta_{23}^{2}}+\theta_{24}^{2}+\theta_{34}^{2})\sin^{2}{(\Delta_{31}+ {\delta c}_{31}L/2)}\\
&\quad-
(c_{12}^{2}\Delta_{12}+2\theta_{24}\theta_{34}\cos{\delta_{3}}\Delta_{n})\sin{2(\Delta_{31}+ {\delta c}_{31}L/2)} -\dfrac{s_{13}^{2} (\Delta_{31}+ {\delta c}_{31}L/2)}{(\Delta_{31}+ {\delta c}_{31}L/2-\Delta_{e})^{2}}\\
&\quad
[2(\Delta_{31}+ {\delta c}_{31}L/2)\sin{(\Delta_{31}+ {\delta c}_{31}L/2)}\cos{\Delta_{e}}
\sin{(\Delta_{31}+ {\delta c}_{31}L/2-\Delta_{e})}\\
&\quad -
(\Delta_{31}+ {\delta c}_{31}L/2-\Delta_{e})\Delta_{e}\sin{2(\Delta_{31}+ {\delta c}_{31}L/2)}]\\
&\quad+ \theta_{24}\theta_{34}\sin{\delta_{3}}\sin{2(\Delta_{31}+ {\delta c}_{31}L/2)}
\end{split}
\end{equation}
\end{center}
\vspace{-4ex}
\begin{center}
\begin{equation}
 P_{\mu s}=2\theta_{24}^{2}+(\theta_{34}^{2}-\theta_{24}^{2})\sin^{2}({\Delta_{31}+ {\delta c}_{31}L/2})-\theta_{24}\theta_{34}\sin{\delta_{3}}\sin{2(\Delta_{31}+ {\delta c}_{31}L/2)} 
\end{equation}
\end{center} 
For small angles ($\theta_{ij}\simeq \sin{\theta_{ij}}\simeq s_{ij}$) these oscillation probabilities can be written as  \vspace{-4ex}
\begin{center}
\begin{equation}
P_{ee} =1-2{s_{14}}^{2}-4{s_{13}}^{2}(\Delta_{31}+ {\delta c}_{31}L/2)^{2}\dfrac{\sin^{2}{(\Delta_{31}+ {\delta c}_{31}L/2-\Delta_{e})}}{(\Delta_{31}+ {\delta c}_{31}L/2-\Delta_{e})^{2}}
\end{equation}
\end{center}
\vspace{-4ex}
\begin{center}
\begin{equation}
P_{e\mu}=P_{e\tau}=2 s_{13}^{2}(\Delta_{31}+ {\delta c}_{31}L/2)^{2}\dfrac{\sin^{2}{(\Delta_{31}+ {\delta c}_{31}L/2-\Delta_{e})}}{(\Delta_{31}+ {\delta c}_{31}L/2-\Delta_{e})^{2}}
\end{equation}
\end{center}
\vspace{-4ex}
\begin{center}
\begin{equation}
\begin{split}
P_{\mu \mu} & =1-2s_{24}^{2}\cos^{2}{(\Delta_{31}+ {\delta c}_{31}L/2)}-(1-8\hat{s_{23}^{2}})\sin^{2}{(\Delta_{31}+ {\delta c}_{31}L/2)}
\\
&\quad +(c_{12}^{2}
\Delta_{12}-2s_{24}s_{34}\cos{\delta_{3}}\Delta_{n})\sin{2(\Delta_{31}+ {\delta c}_{31}L/2)}
+\dfrac{s_{13}^{2}(\Delta_{31}+ {\delta c}_{31}L/2)}{(\Delta_{31}+ {\delta c}_{31}L/2-\Delta_{e})^{2}} \\
&\quad \times [2(\Delta_{31}+ {\delta c}_{31}L/2)\sin{\Delta_{e}}\cos{(\Delta_{31}+ {\delta c}_{31}L/2)}
\sin{(\Delta_{31}+ {\delta c}_{31}L/2-\Delta_{e})}- \\
&\quad
(\Delta_{31}+ {\delta c}_{31}L/2-\Delta_{e})\Delta_{e}\sin{2(\Delta_{31}+ {\delta c}_{31}L/2)} ]
\end{split}
\end{equation}
\end{center}
\vspace{-4ex}
\begin{center}
\begin{equation}
\begin{split}
P_{\mu\tau} & = \sin^{2}{(\Delta_{31}+ {\delta c}_{31}L/2)}-(8\hat{s_{23}^{2}+s_{24}^{2}+s_{34}^{2}})\sin^{2}{(\Delta_{31}+ {\delta c}_{31}L/2)}-\\
&\quad (c_{12}^{2}\Delta_{12}+2s_{24}s_{34}\cos{\delta_{3}}\Delta_{n})\sin{2(\Delta_{31}+ {\delta c}_{31}L/2)}-\dfrac{s_{13}^{2}(\Delta_{31}+ {\delta c}_{31}L/2)}{(\Delta_{31}+ {\delta c}_{31}L/2-\Delta_{e})^{2}} \\
&\quad \times [2(\Delta_{31}+ {\delta c}_{31}L/2)\sin{(\Delta_{31}+ {\delta c}_{31}L/2)}\cos{\Delta_{e}}
\sin{( \Delta_{31}+ {\delta c}_{31}L/2-\Delta_{e})}-\\
&\quad
(\Delta_{31}+ {\delta c}_{31}L/2-\Delta_{e})\Delta_{e}\sin{2(\Delta_{31}+ {\delta c}_{31}L/2)}]+ s_{24}s_{34}\sin{\delta_{3}}\sin{2(\Delta_{31}+ {\delta c}_{31}L/2)}
\end{split}
\end{equation}
\end{center}
\vspace{-4ex}
\begin{center}
\begin{equation}
 P_{\mu s}=2s_{24}^{2}+(s_{34}^{2}-s_{24}^{2})\sin^{2}{(\Delta_{31}+ {\delta c}_{31}L/2)}-s_{24}s_{34}\sin{\delta_{3}}\sin{2(\Delta_{31}+ {\delta c}_{31}L/2)} 
\end{equation}
\end{center}
In order to analyse CPT violation at probability level in 4 flavor neutrino framework the value of $\Delta P^{CPT}_{\alpha\beta}$ considered in our analysis is given by    \vspace{-6ex}
\begin{center}
\begin{equation}
 \bigtriangleup P_{\alpha \beta}^{CPT}=[(P_{\alpha \beta})_{4\nu}]_{\delta c_{ij}\neq 0} - [(P_{\alpha \beta})_{4\nu}]_{\delta c_{ij} = 0}
\end{equation}
\end{center}
\vspace{-4ex}
\FloatBarrier
\begin{figure}[htbp]
\begin{minipage}[b]{0.53\linewidth}
\centering
\includegraphics[width=\textwidth]{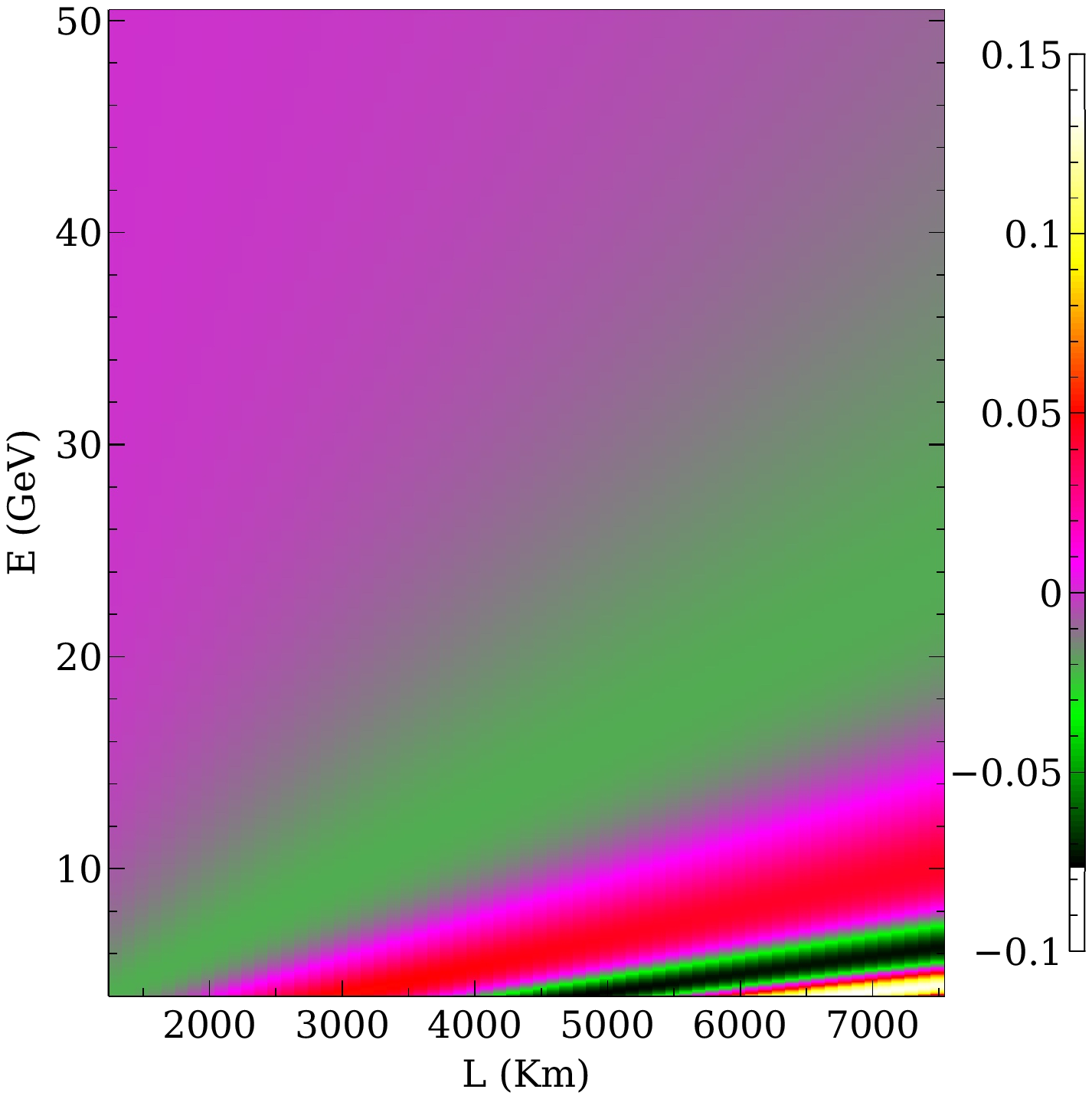}
\subfigure{(a)}
\end{minipage}
\begin{minipage}[b]{0.53\linewidth}
\centering
\includegraphics[width=\textwidth]{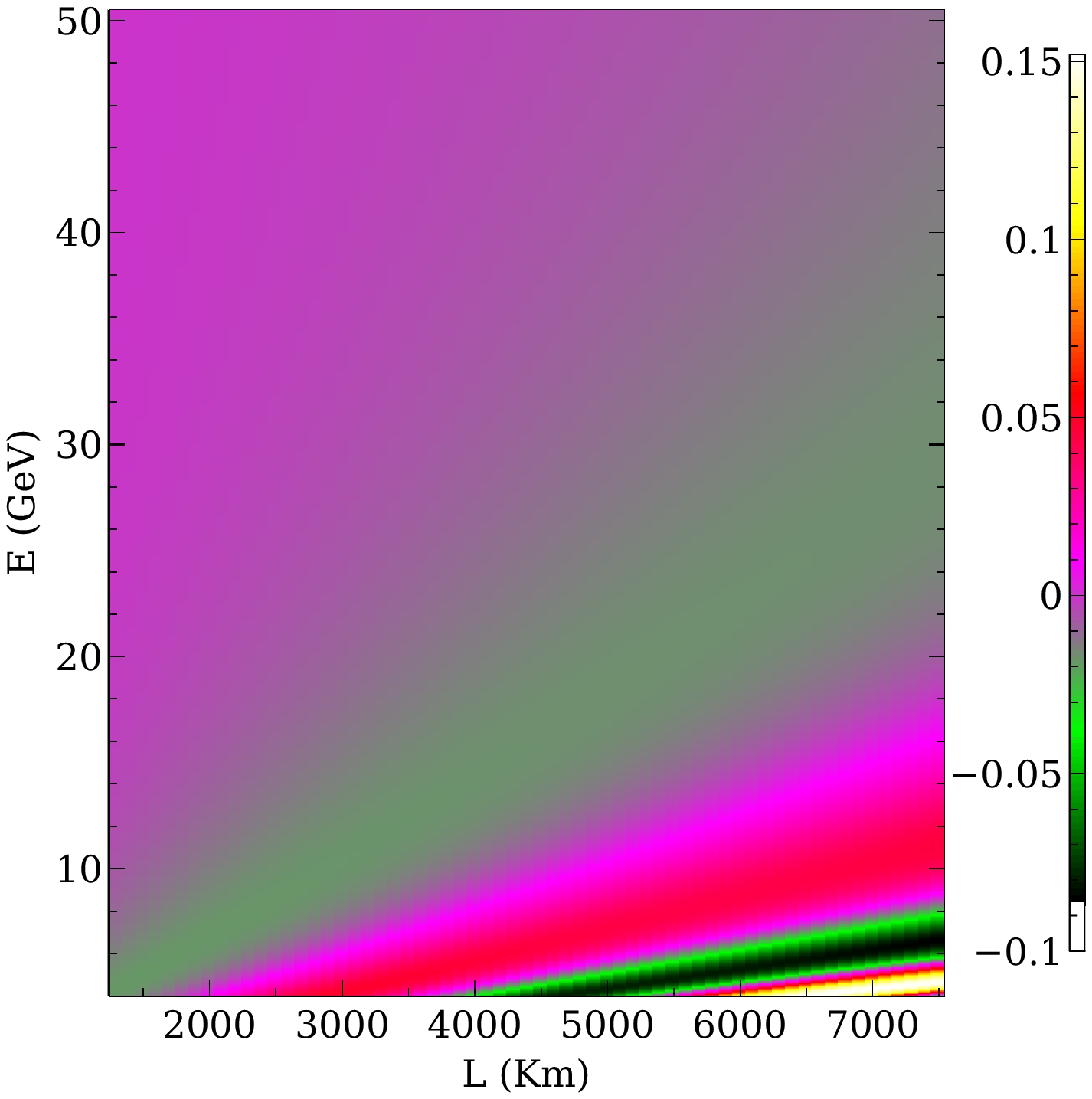}
\subfigure{(b)}
\end{minipage}
\end{figure}
\FloatBarrier
\begin{figure}[htbp]
\begin{minipage}[b]{0.53\linewidth}
\centering
\includegraphics[width=\textwidth]{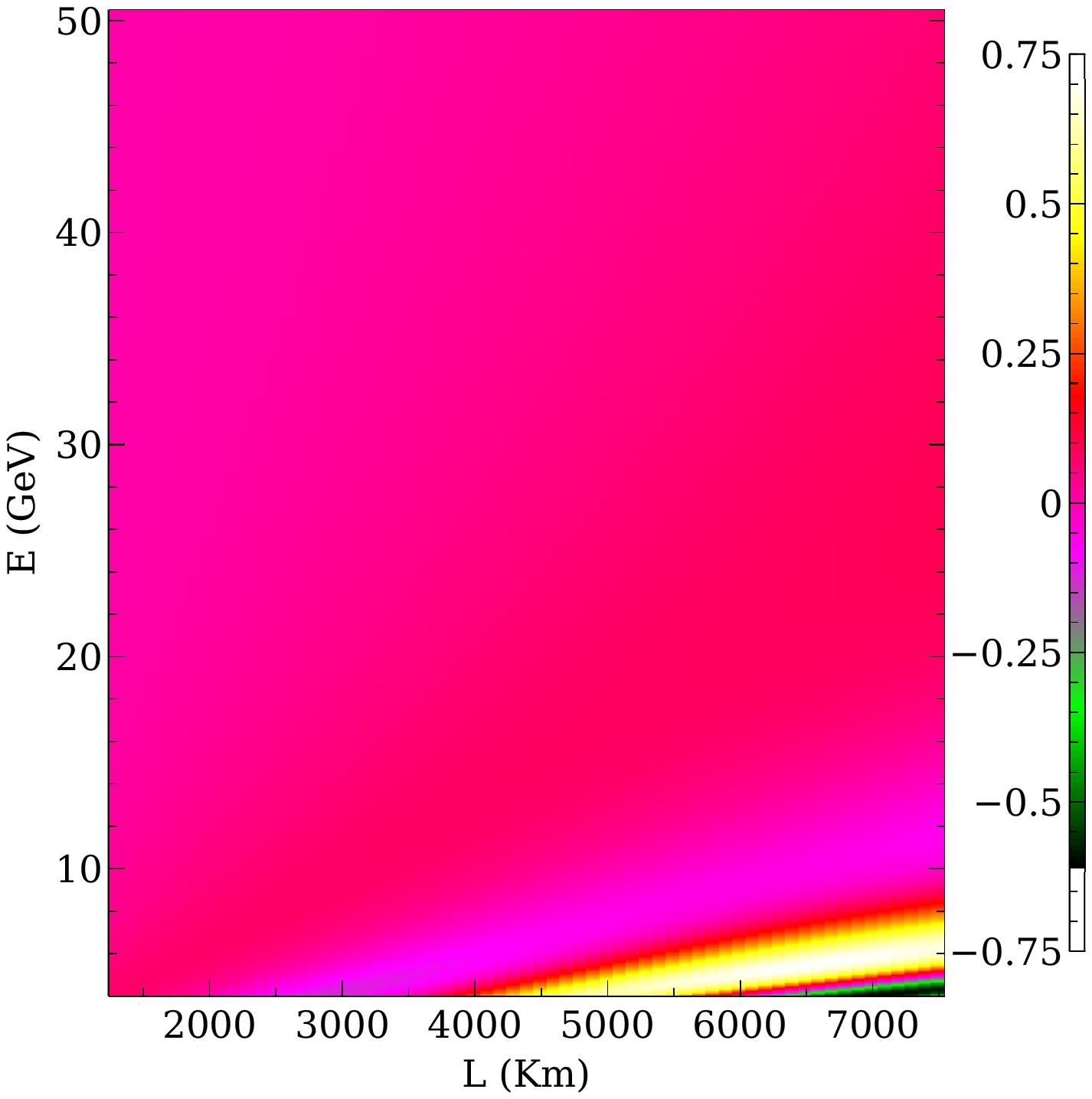}
\subfigure{(c)}
\end{minipage}
\hfill
\begin{minipage}[b]{0.53\linewidth}
\centering
\includegraphics[width=\textwidth]{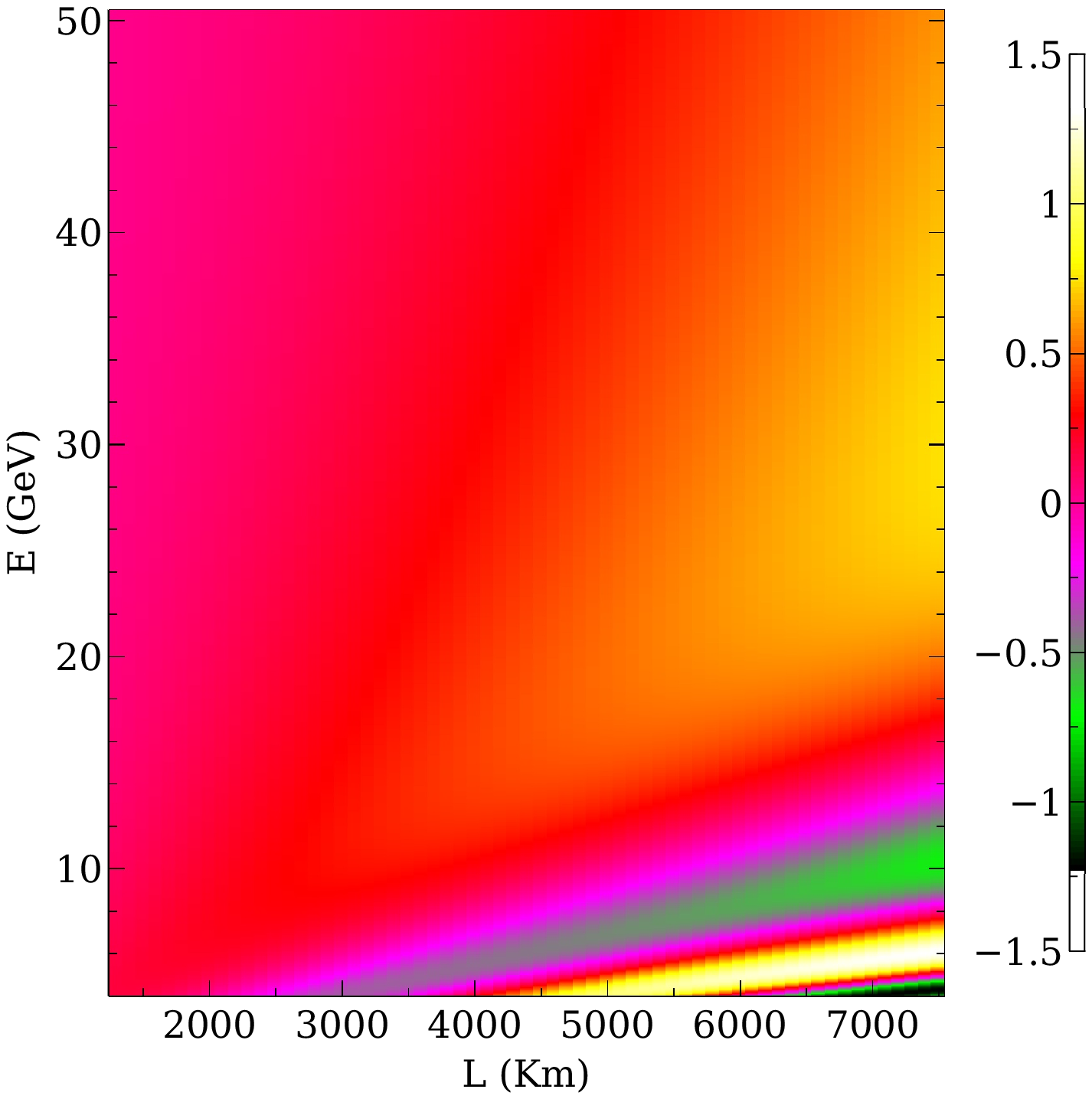}
\subfigure{(d)}
\end{minipage}
\caption{The  oscillographs (a) and (b) demonstrates the variation of $ \Delta P^{CPT}_{\mu\mu} $ with  baseline and energy for normal hierarchy whereas (c) and (d) illustrates the same for inverted hierarchy. For the oscillographs (a) and (c) the values of sterile parameters are selected from long baseline experiments while for (b) and (d) the values of sterile parameters are taken from reactor and atmospheric experiments.}
\end{figure}

\FloatBarrier
\begin{figure}[htbp]
\begin{minipage}[b]{0.55\linewidth}
\centering
\includegraphics[width=\textwidth]{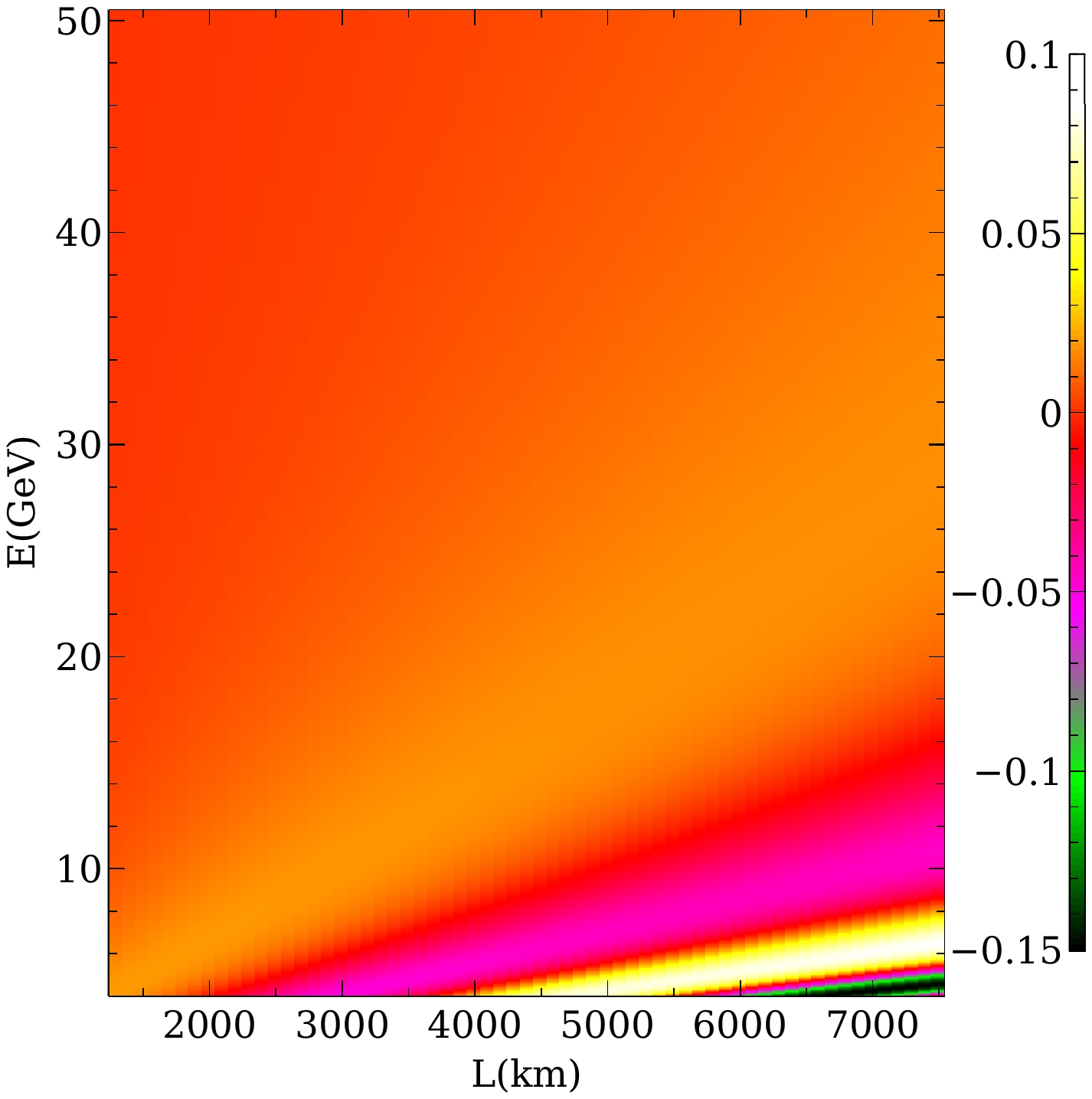}
\subfigure{(a) Normal hierarchy}
\end{minipage}
\hspace{0.0cm}
\begin{minipage}[b]{0.55\linewidth}
\centering
\includegraphics[width=\textwidth]{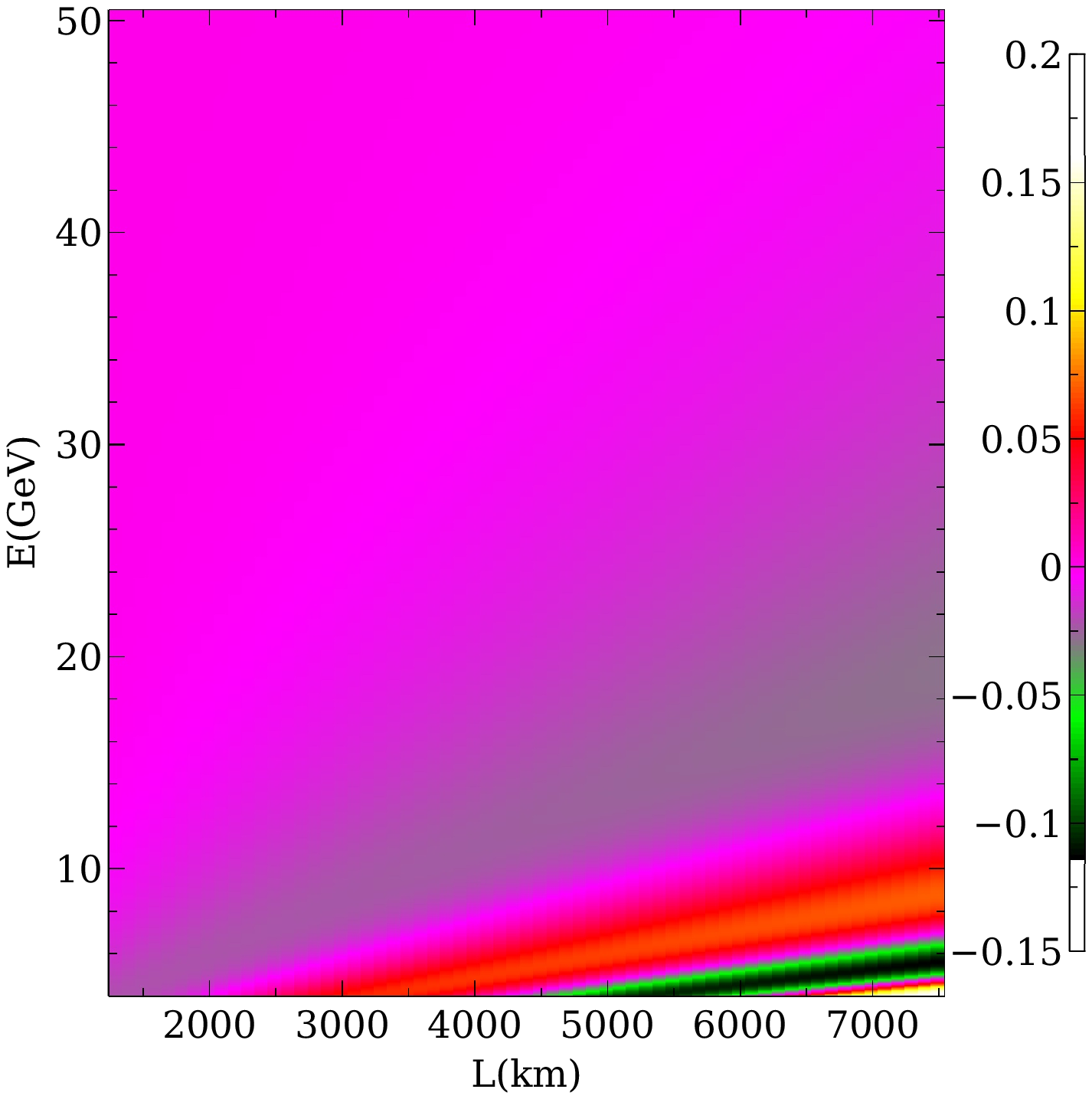}
\subfigure{(b) Inverted hierarchy }
\end{minipage}
\caption{The oscillographs demonstrate the variation of $ \Delta P^{CPT}_{\bar{\mu}\bar{\mu}} $ with  baseline and energy. The values of sterile parameters are selected from reactor+atm experiments. Left  and right oscillographs are for normal and inverted hierarchies respectively.}
\end{figure}

Neutrino factory setup considered for analysing CPT violation is taken from the references \cite{International}\cite{rolinec}\cite{winter1}\cite{kopp1} and \cite{Albright1}. The experimental setup and detector specifications considered in our analysis are mentioned below. Neutrino factory setup consist of $1.4 \times 10^{21}$ useful muon decays per polarity, with  parent muon energy $E_{\mu}=50 GeV$ . We have done our analysis for 10 years running of neutrino factory. In particle physics meaningful observations always demands a detector with very good energy and angular resolutions. This view point lead us to select Liquid Argon detector for particle detection. The energy resolution of the detector for muon  is $\sigma(GeV) = 0.20/ \sqrt{E_{\nu}(GeV)}$.  \\
A near detector is placed at a distance 20 m from the end of the decay straight of the muon storage ring. Effective baseline ($L_{eff}$) is used in place of baseline (L), which is calculated using  $ L_{eff}=\sqrt{d(d+s)}  $ \cite{winter}. Fiducial mass of near detector is 200 tons. Presence of near detector will minimize the systematic uncertainties in our observations.\\
A 50 Kt far detector is placed  at a distance of 7500 Km. The systematic uncertainties considered for this analysis are given in Table 1. An uncertainty  of  5\%  on  matter density \cite{geller}\cite{ohlsson} is also considered in our work . The simulated environment of the neutrino factory is created with the help of GLoBES \cite{Huber1} \cite{Huber2}. The analytical equations for four flavor neutrino conversion probabilities derived in this work are defined in the probability engine of the software. The best fit values of oscillation parameters \cite{meloni} \cite{garcia} are mentioned in Table 2. Sterile parameter values mentioned in the Table 2 represents best fit values for $ \Delta m_{41}^{2} $ =0.1 $ eV^{2} $. 
\begin{table}[ht]
\caption{Systematics }
% title of Table
\centering
% used for centering table
\begin{tabular}{c c c c}
% centered columns (4 columns)
\hline\hline
%inserts double horizontal lines

Systematic uncertainties  &  values  \\ [0.5ex]
% inserts table
%heading
\hline
% inserts single horizontal line
Flux normalization & 2\% \\
% inserting body of the table
Fiducial mass errors for near detector & 0.6 \%\\
Fiducial mass errors for far detector  & 0.6 \%\\
energy calibration error for near detector & 0.5 \%\\
energy calibration error for far detector  & 0.5\\ 
shape error & 10 \%\\
Backgrounds &  $10^{-4}$       \\

% [1ex] adds vertical space
\hline
%inserts single line
\end{tabular}
\label{table:nonlin}
% is used to refer this table in the text
\end{table}

\begin{table}[ht]
\caption{Best fit values of the oscillation parameters}
% title of Table
\centering
% used for centering table
\begin{tabular}{c c c c}
% centered columns (4 columns)
\hline\hline
%inserts double horizontal lines
Parameter & Best fit values  \\ [0.5ex]
% inserts table
%heading
\hline
% inserts single horizontal line
$\theta_{12}$ & $34.4^\circ$  \\
% inserting body of the table
$\theta_{13}$ & $8.50^\circ$ \\
$\theta_{23}$ & $45.0^\circ$ \\
$\theta_{14}$ & $6.7^\circ$; $10^\circ$ \\
$\theta_{24}$ & $3.3^\circ$ ; $12^\circ$\\ 
$\theta_{34}$ & $6.3^\circ$ ; $28^\circ$ \\
${\Delta m_{21}}^{2}$ & $8\times 10^{-5} eV^2$ \\
${\Delta m_{31}}^{2}$ & $2.5\times 10^{-3} eV^2$ \\[1ex]
% [1ex] adds vertical space
\hline
%inserts single line
\end{tabular}
\label{table:nonlin}
% is used to refer this table in the text
\end{table}
The constraints on CPT violating parameters  $ \delta c_{21} $ and $ \delta c_{31} $ within two and three neutrino frameworks are mentioned in references  
\cite{animesh}  \cite{anindya} \cite{Bahcall}  \cite{Barger1} \cite{Dighe} \cite{Samanta} and \cite{Gonzalez}. In present work we are trying to check the neutrino factory potential to capture CPT violating signatures in presence of sterile neutrino. As the mass hierarchy determination is yet in research phase therefore in an attempt to make this work relevant we have analysed CPT invariance for both the mass hierarchies. Initially CPT violating signatures are checked at probability level. The value of  $\Delta P^{CPT}_{\alpha\beta}$ is estimated  by substituting equation (41) in equation (44) for the channel $ \nu_{\mu} \rightarrow \nu_{\mu} $.  The variation in  $\Delta P^{CPT}_{\mu \mu}$ and  $\Delta P^{CPT}_{\bar{\mu}\bar{\mu}}$ with baseline and energy is shown in Figure 2 and Figure 3 oscillographs respectively. The total CPT violation captured by any experiment will be the sum of extrinsic CPT violation (CPT violation arising due to matter effects) and intrinsic or genuine CPT violation(which we are probing in present work). In an endeavour to constraint intrinsic CPT violating parameters we must look for places where extrinsic CPT violation is negligible or very less. With three active neutrinos the extrinsic CPT violation is checked in reference \cite{magnus} whereas with 3 (active) + 1 (sterile) neutrinos it is checked in reference \cite{sujata}. These references conclude that extrinsic CPT violation for energies 4 GeV- 6 GeV is negligible at shorter baselines, roughly less than 2000 km. Whereas for longer baselines this effect decreases with energy. Equation (44) of our work will check the presence of pure CPT violation arising in the presence of sterile neutrino at probability level. The values of CPT violating parameters considered while plotting oscillographs are $ \delta c_{31}=4\times10^{-23} $ GeV  and $ \delta c_{21}=3\times10^{-23} $ GeV. Looking at normal and inverted hierarchy oscillographs(Figure 2) we can observe the presence of pure CPT violating signatures at shorter baselines i.e. from 1300 km-2000 km for 4 GeV-6 GeV energies. The references \cite{magnus} \cite{sujata}, which speak about extrinsic CPT violation have recorded very weak or almost negligible signatures of extrinsic CPT violation at the above mentioned energies and baselines. Hence baselines from 1300 km-2000 km with neutrino energies in the range 4 GeV-6 GeV are favourable for probing CPT violation with neutrino factory. In Figure 2 while looking at normal hierarchy oscillographs we observe $\Delta P^{CPT}_{\mu \mu}$ =-0.05 along baselines 4000 km-7500 km for energies 12 GeV-30 GeV. The inverted hierarchy oscillographs of the same Figure captures $\Delta P^{CPT}_{\mu \mu}$=0.25 and 1 for sterile parameters taken from long baseline experiments and reactor+atmospheric experiments respectively. This probability difference can be observed for baselines 4000 km-7500 km and energies 20 GeV-50 GeV.\\ 
After examining the presence of pure CPT violation at probability level we go ahead to observe the signatures of the same with realistic proposed neutrino experiments i.e. neutrino factory. The specifications of neutrino factory considered in our work are mentioned earlier. Liquid argon detector seems a reasonable choice to grab the signatures of leptons in the considered energy range. The rate(event) level analysis depends on mathematical formulation( oscillation probability), physics(types of interactions) and R \& D (source properties and detector properties) of the experiment. Looking at equations (34) to (44) we found that $ \delta c_{21} $  appears with  $ \Delta m_{21}^{2}  $ term and $ \delta c_{31} $ term appears with $ \Delta m_{31}^{2}$ term. The solar and atmospheric mass square difference ($ \Delta m_{21}^{2}  $ and $ \Delta m_{31}^{2}  $) are of the order of $ 10^{-23} $ and $ 10^{-21} $ respectively. Hence, any change in mass terms due to the presence of CPT violating parameter will be better observed in $ \delta c_{31}$ term.\\
To hook CPT violating impression with neutrino factory we have investigated some observable parameters like R, $ \bigtriangleup R $ and  asymmetry factor. These terms are defined by equations (45) and (46). The ratio R and ratio difference $ \Delta R $ are examined as
 \vspace{-8ex} \begin{center}
\begin{equation}
R =\frac{N(\nu_{\mu}\rightarrow \nu_{\mu})}{N(\bar{\nu_{\mu}}\rightarrow \bar{\nu_{\mu}})} ;
\bigtriangleup R =(R_{4\nu})_{\delta c_{ij}\neq 0} - (R_{4\nu})_{\delta c_{ij} = 0} 
\end{equation}
\end{center}  \vspace{-2ex} 
where $N(\nu_{\mu}\rightarrow \nu_{\mu})$ denotes number of muon neutrinos reaching at detector as muon neutrinos and producing a $ \mu^{-} $ lepton and $N(\bar{\nu_{\mu}}\rightarrow \bar{\nu_{\mu}})$ denotes  number of anti muon neutrinos reaching at detector as anti muon neutrinos and producing a $ \mu^{+} $ lepton. In  $ \bigtriangleup R $, $ (R_{4\nu})_{\delta c_{ij}\neq 0} $  denotes the ratio R in presence of CPT violating terms and $ (R_{4\nu})_{\delta c_{ij} = 0} $ denotes the ratio R in absence of CPT violating terms.
 \FloatBarrier
\begin{figure}[htbp]
\begin{minipage}[b]{0.55\linewidth}
\centering
\includegraphics[width=\textwidth]{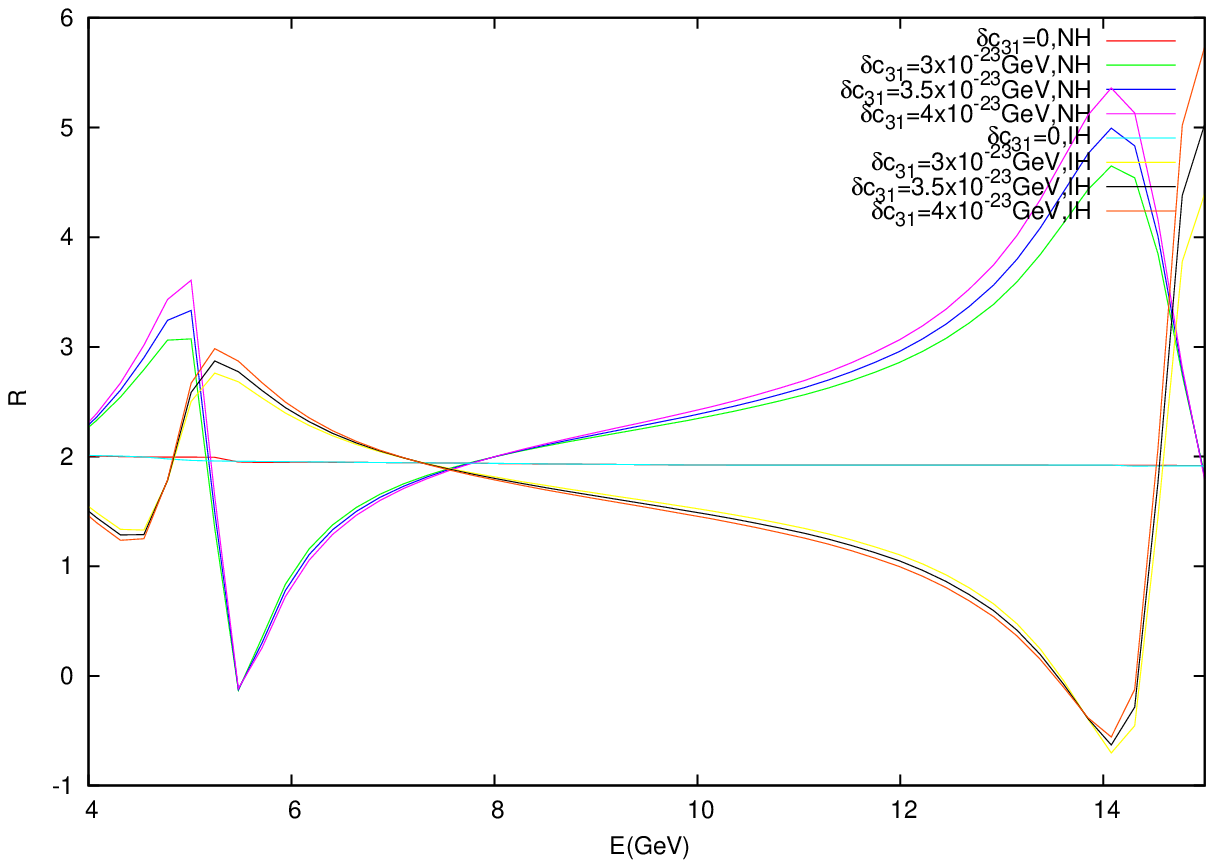}
\subfigure{(a)}
\end{minipage}
\hspace{0.0cm}
\begin{minipage}[b]{0.55\linewidth}
\centering
\includegraphics[width=\textwidth]{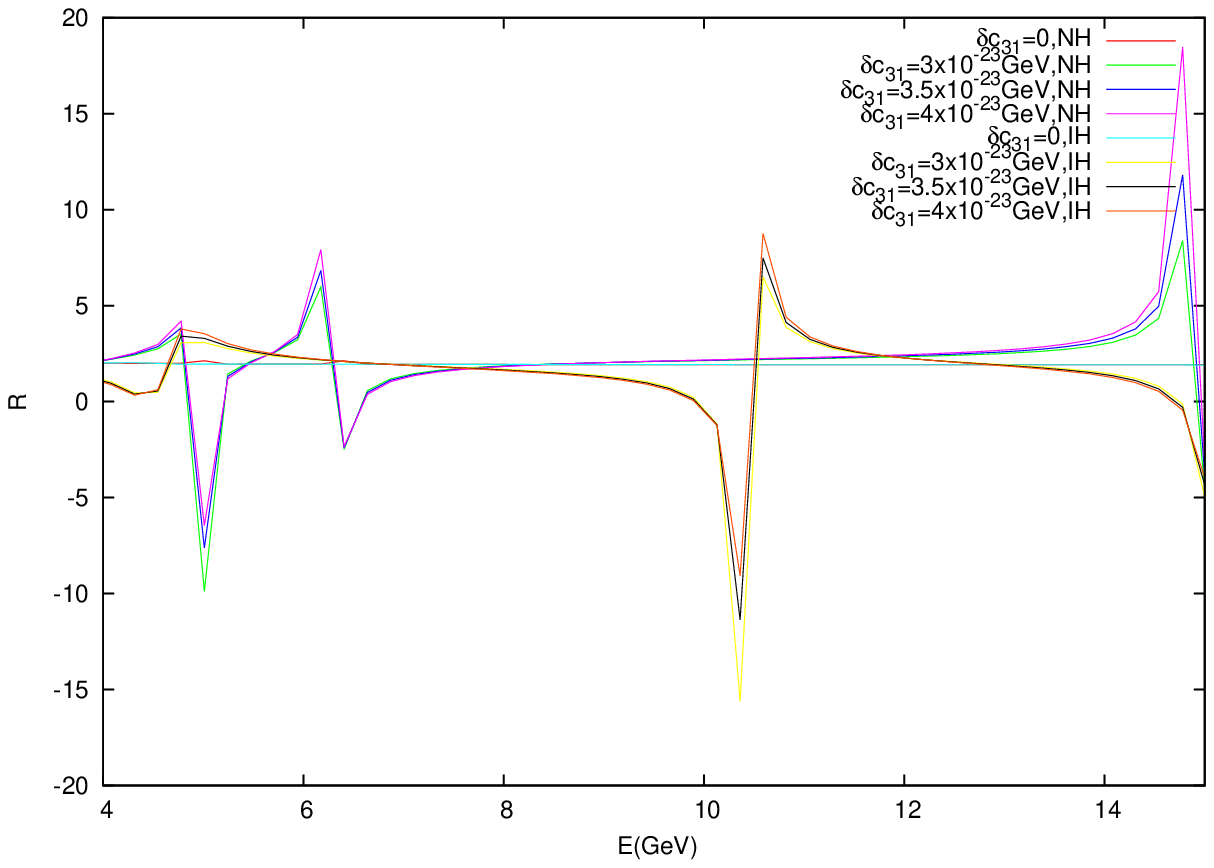}
\subfigure{(b)}
\end{minipage}
\caption{ The  variation of R with energy in the energy range 4-15 GeV. Values of sterile parameters considered in plot (a) and plot (b) are selected from long baseline experiments and  reactor+atm experiments respectively. These observations are made for different values of CPT violating parameter $ \delta c_{31}$;
(i) $ \delta c_{31}=0 $ (setting CPT violating parameter to zero)
(ii) $ \delta c_{31}=3\times10^{-23} $ GeV
(iii) $ \delta c_{31}=3.5\times10^{-23} $ GeV                
(iv) $ \delta c_{31}=4\times10^{-23} $ GeV.
For all the observations $ \delta c_{21}=3\times10^{-23} $ GeV
}
\end{figure}
\FloatBarrier
\begin{figure}[htbp]
\begin{minipage}[b]{0.55\linewidth}
\centering
\includegraphics[width=\textwidth]{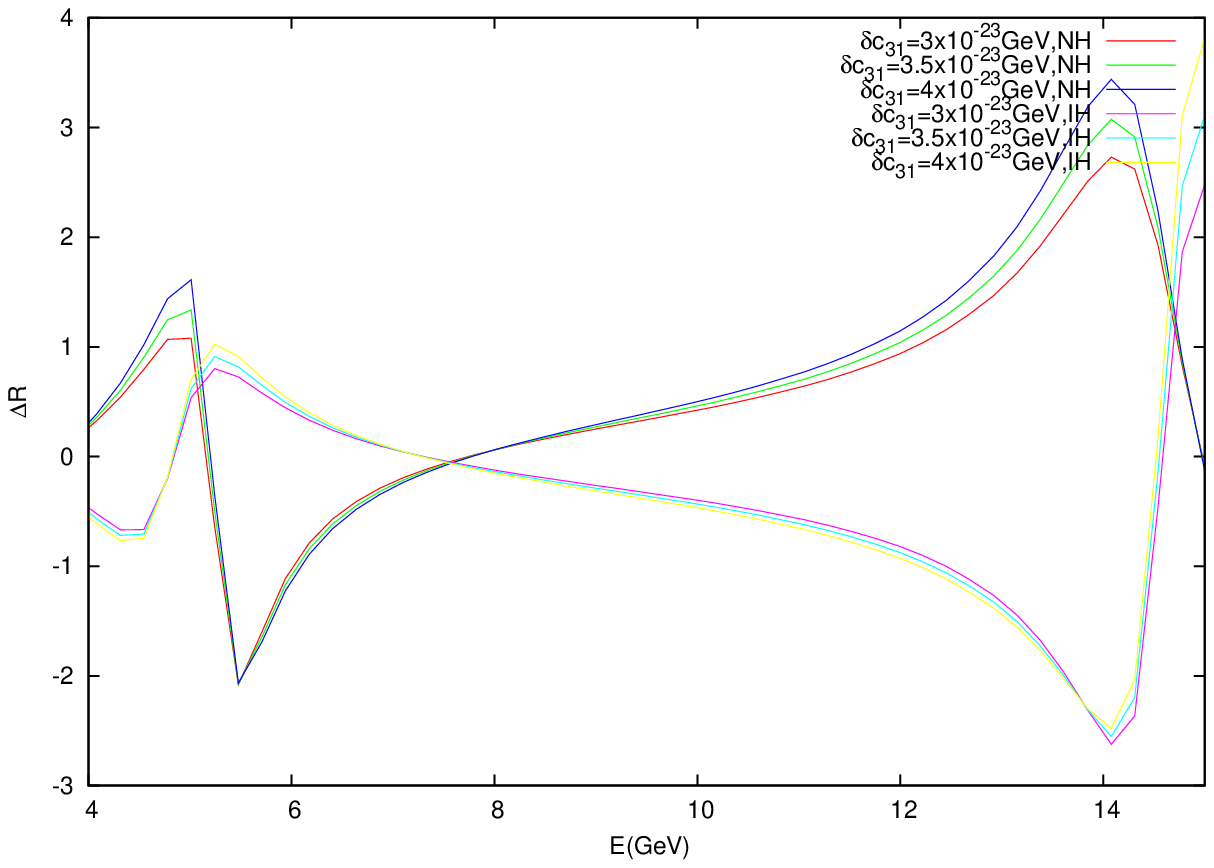}
\subfigure{(a)}
\end{minipage}
\hspace{0.0cm}
\begin{minipage}[b]{0.55\linewidth}
\centering
\includegraphics[width=\textwidth]{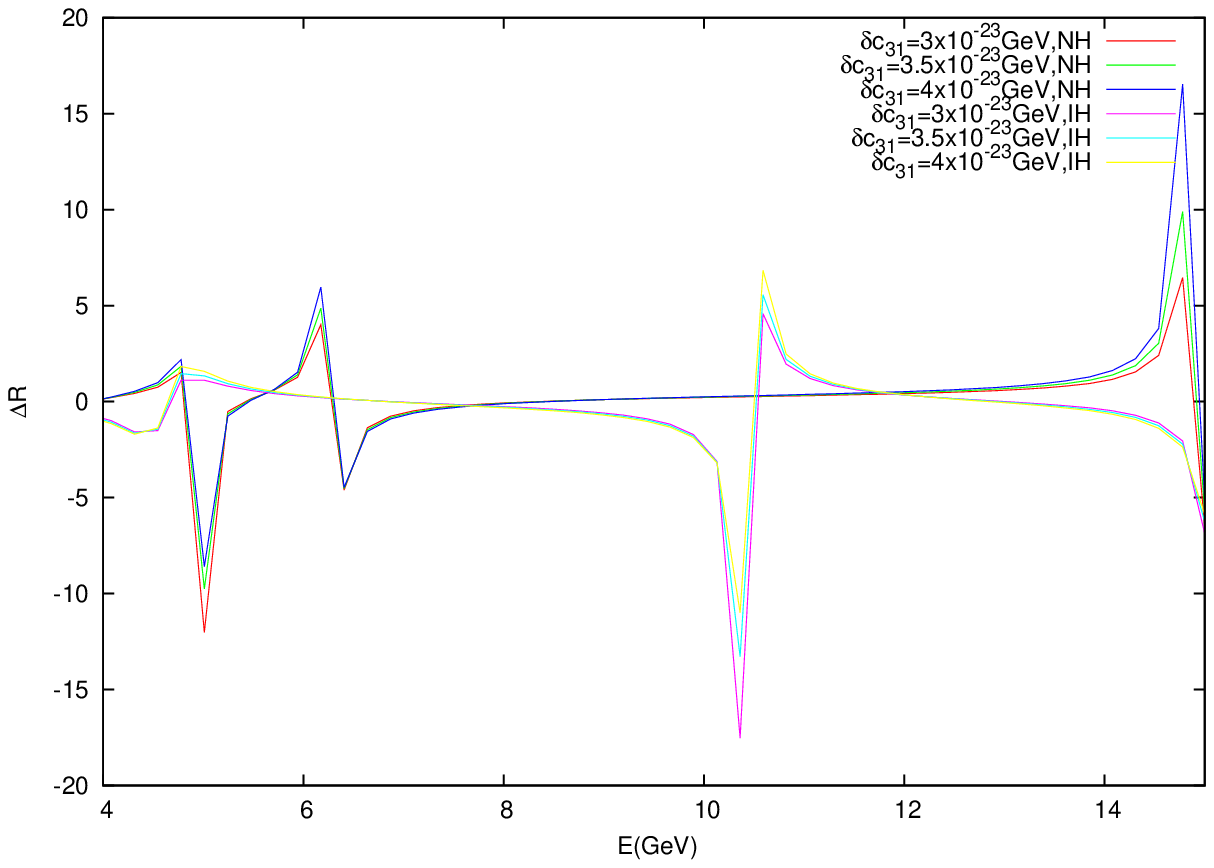}
\subfigure{(b)}
\end{minipage}
\caption{ The  variation of $ \bigtriangleup R $ with energy in the energy range 4-15 GeV. Values of sterile parameters considered in plot (a) and plot (b) are selected from long baseline experiments and  reactor+atm experiments respectively. These observations are made for different values of CPT violating  parameter $ \delta c_{31}$;
(i) $ \delta c_{31}=0 $ (setting CPT violating parameter to zero)
(ii) $ \delta c_{31}=3\times10^{-23} $ GeV
(iii) $ \delta c_{31}=3.5\times10^{-23} $ GeV                
(iv) $ \delta c_{31}=4\times10^{-23} $ GeV.
For all the observations $ \delta c_{21}=3\times10^{-23} $ GeV
}
\end{figure}
\FloatBarrier
\begin{figure}[htbp]
\begin{minipage}[b]{0.55\linewidth}
\centering
\includegraphics[width=\textwidth]{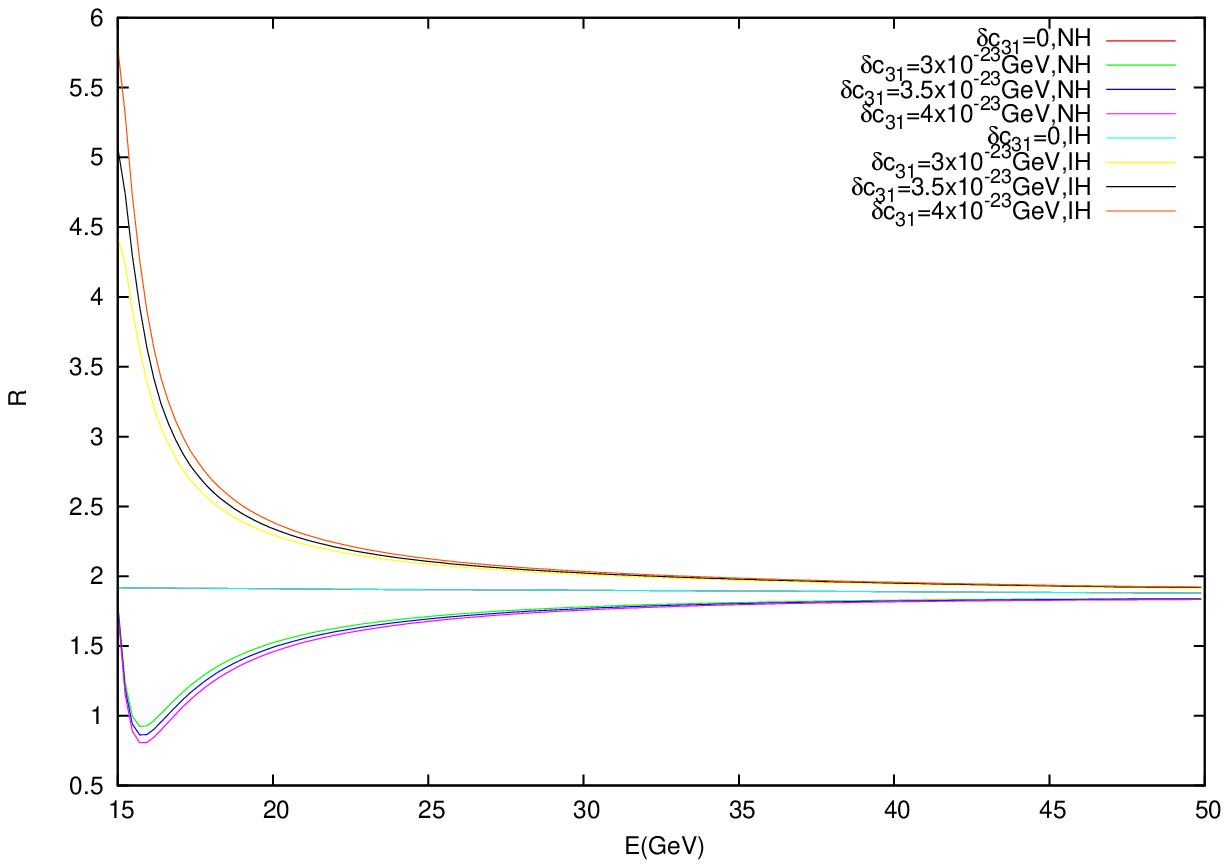}
\subfigure{(a)}
\end{minipage}
\hspace{0.0cm}
\begin{minipage}[b]{0.55\linewidth}
\centering
\includegraphics[width=\textwidth]{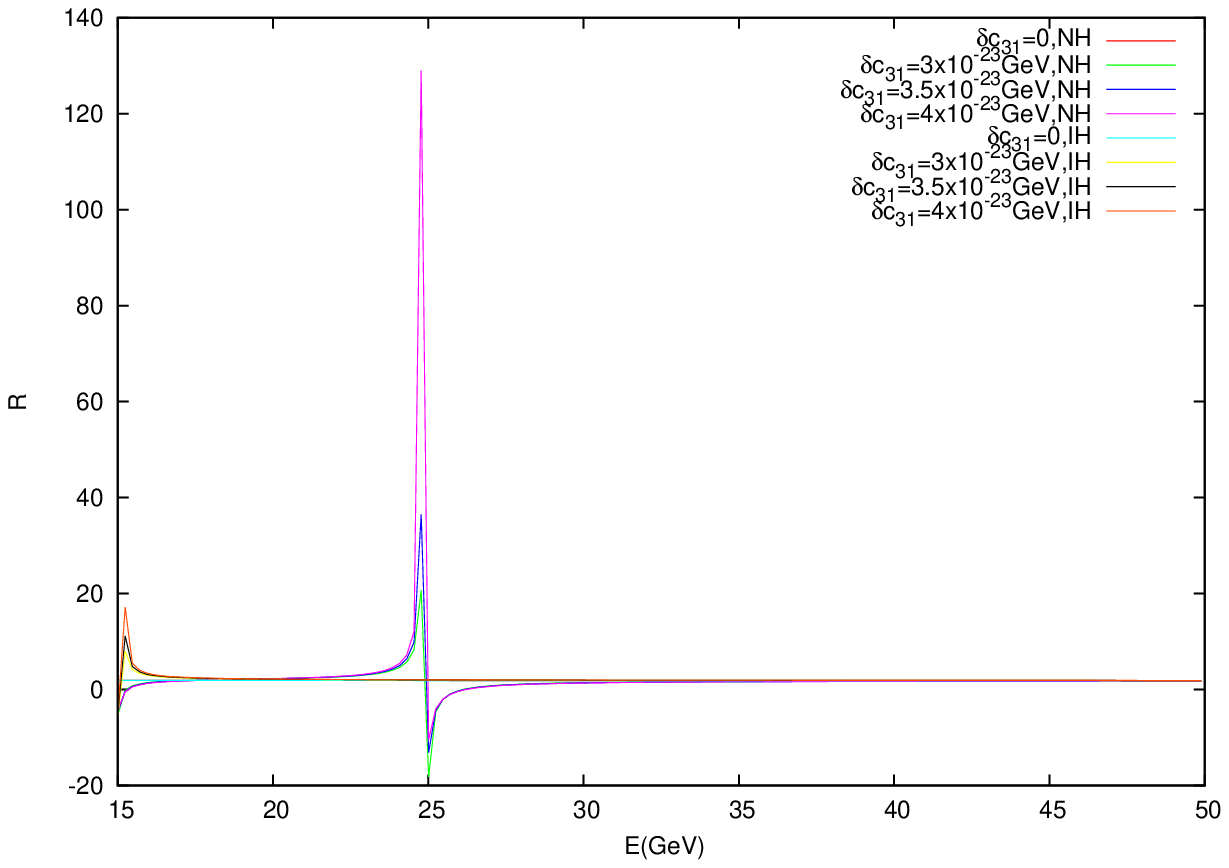}
\subfigure{(b)}
\end{minipage}

\caption{ The  variation of R with energy in the energy range 15-50 GeV. Values of sterile parameters considered in plot(a) and plot(b) are selected from long baseline experiments and reactor+atm experiments respectively. These observations are made for different values of CPT violating parameter $ \delta c_{31}$
(i) $ \delta c_{31}=0 $ (setting CPT violating parameter to zero)
(ii) $ \delta c_{31}=3\times10^{-23} $ GeV
(iii) $ \delta c_{31}=3.5\times10^{-23} $ GeV                
(iv) $ \delta c_{31}=4\times10^{-23} $ GeV.
For all the observations $ \delta c_{21}=3\times10^{-23} $ GeV
}
\end{figure}
\FloatBarrier
\begin{figure}[htbp]
\begin{minipage}[b]{0.55\linewidth}
\centering
\includegraphics[width=\textwidth]{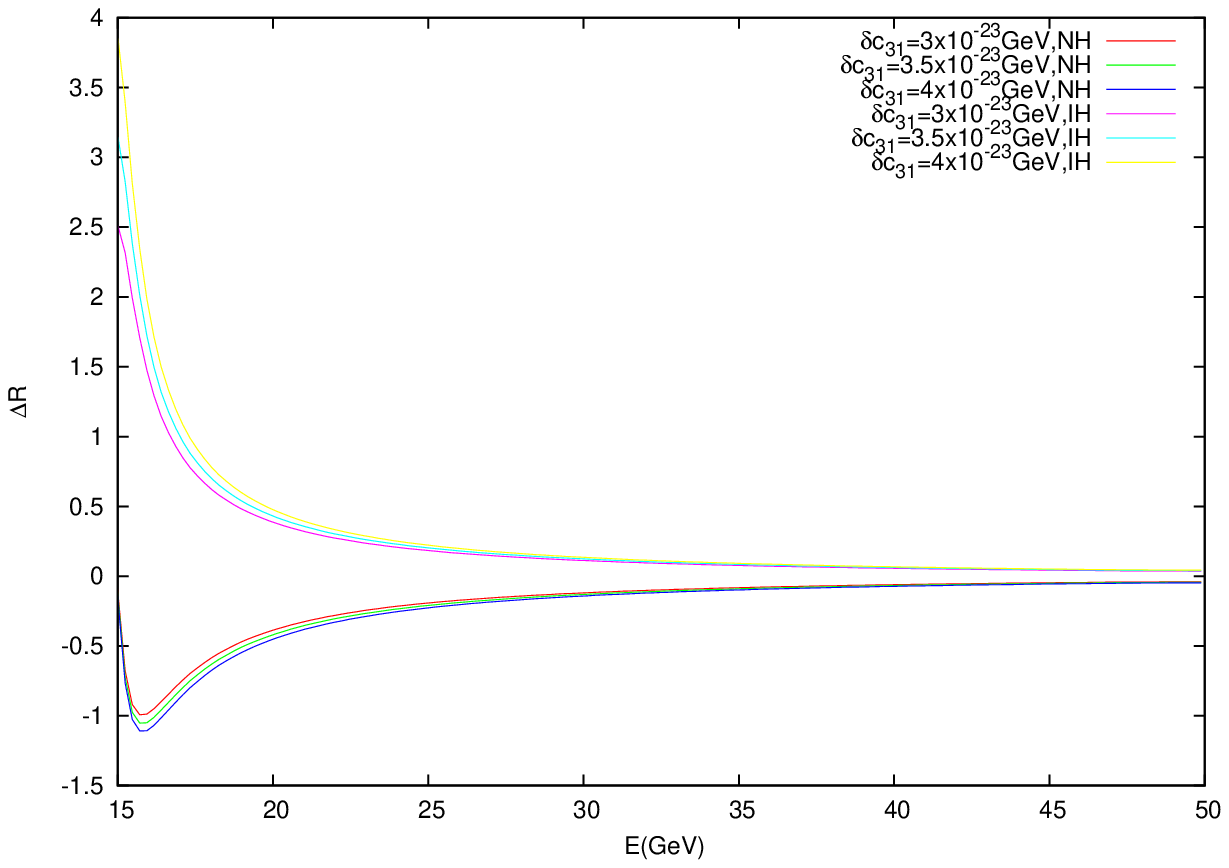}
\subfigure{(a)}
\end{minipage}
\hspace{0.0cm}
\begin{minipage}[b]{0.55\linewidth}
\centering
\includegraphics[width=\textwidth]{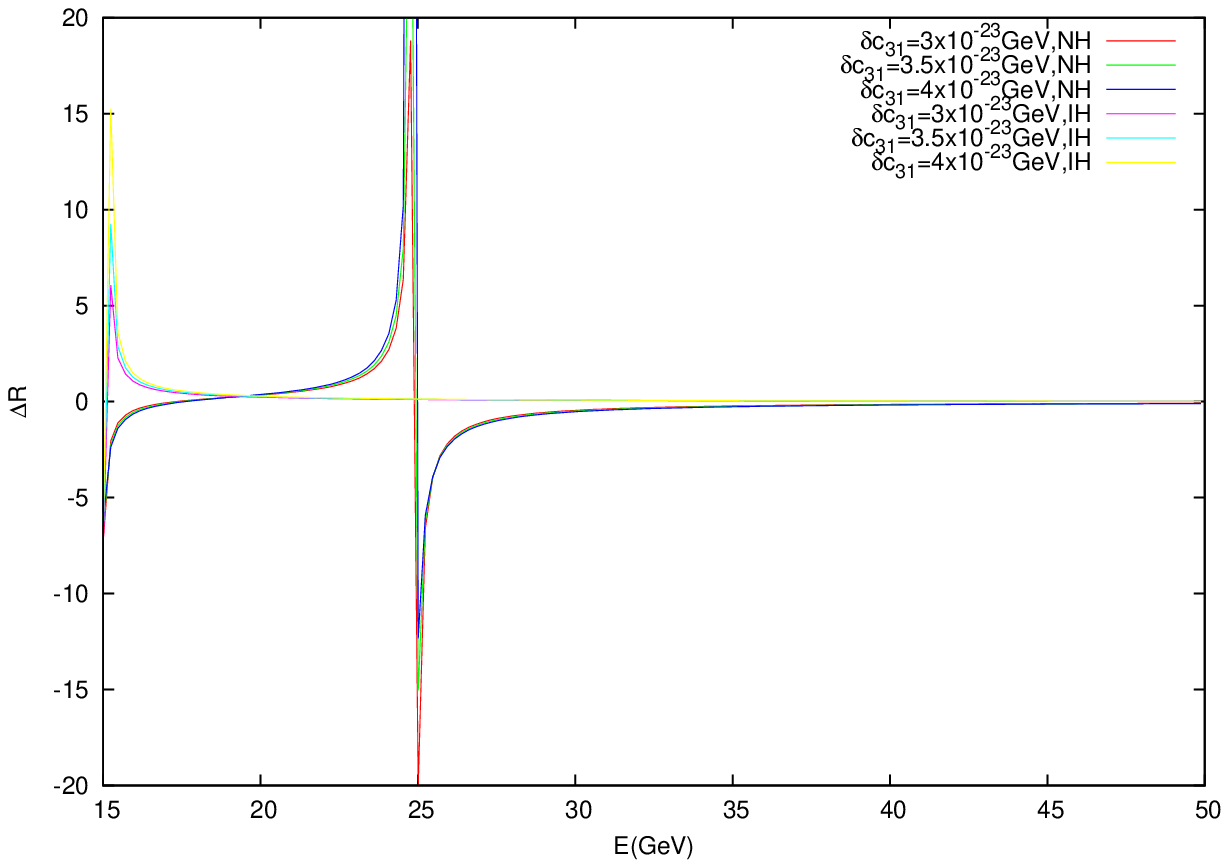}
\subfigure{(b) }
\end{minipage}

\caption{ The variation of $ \bigtriangleup R $ with energy in the energy range 15-50 GeV. Values of sterile parameters considered in plot (a) and plot (b) are selected from long baseline experiments and reactor+atm experiments respectively . These observations are made for different values of  CPT violating parameter $ \delta c_{31}$
(i) $ \delta c_{31}=0 $ (setting CPT violating parameter to zero)
(ii) $ \delta c_{31}=3\times10^{-23} $ GeV
(iii) $ \delta c_{31}=3.5\times10^{-23} $ GeV                
(iv) $ \delta c_{31}=4\times10^{-23} $ GeV.
For all the observations $ \delta c_{21}=3\times10^{-23} $ GeV
}
\end{figure}
In presence of matter the observable R will not be equal to one, even if pure CPT violation is absent. It will be equal to a numerical value representing the ratio of neutrino and antineutrino interaction cross-sections. If we want to analyse the extent of deviation produced by pure CPT violation, we have to hide  or filter out the deviation produced by any other phenomenon. In an attempt to filter out pure CPT violating contribution from the total observed deviation we take into record a new observable $ \Delta R $. This parameter is defined in the equation (45). Figures 4 and 6 demonstrate the variation in R with energy whereas Figures 5 and 7 exhibit variation in $ \Delta R $ with energy for baseline 7500 km. We observe that at long baselines pure  CPT violating effects get smaller with increase in energy. The presence of CPT violation signatures can be observed with neutrino factory and it can be checked by looking R and $ \Delta R $ plots (Figure 4- Figure 7) for different values of CPT violating parameter $ \delta c_{31} $. As we know that in presence of sterile neutrino the manifestation of pure CPT signatures depends on the values of sterile parameters, hence the entire analysis is performed with two sets of best fit values of sterile parameters which were examined by different neutrino experiments. The results from neutrino factory with sterile parameter values obtained from reactor+atmospheric experiments exhibit larger deviation in observables R and $ \Delta R $ in comparison to the results obtained with sterile parameter values taken from long baseline experiments. These observables are checked for both mass hierarchies. From the Figures 4,5,6 and 7 we comprehend that after 15 Gev there is a flip in sign of the observables for both the hierarchies. At the same time the amount of deviation measured for pure CPT violating effects are different for NH and IH for the same energy and baseline.\\  The next observable asymmetry factor $ A_{\mu} $ is defined as \vspace{-4ex} 
\begin{center}
\begin{equation}
A_{\mu}(E)\equiv \frac{N(\nu_{\mu}\rightarrow \nu_{\mu})^{far}(E)}{N(\nu_{\mu}\rightarrow \nu_{\mu})^{near}(E)} - \frac{N(\bar{\nu_{\mu}}\rightarrow \bar{\nu_{\mu}})^{far}(E)}{N(\bar{\nu_{\mu}}\rightarrow \bar{\nu_{\mu}})^{near}(E)}
\end{equation}
\end{center}
This ratio is determined by using far and near detectors. The variations in asymmetry factor with energy for baseline 7500 km are shown by Figures 8 and 9. These figures reflect the variations in observable $ A_{\mu} $ for different values of  $ \delta c_{31} $ (CPT violating parameter) and for both mass hierarchies. The  $ \delta c_{31} $=0 will reflect $ A_{\mu} $ values without any contribution from CPT violating terms. The asymmetry factor increases with the increase in the value of  $ \delta c_{31} $. An enhancement in magnitude of asymmetry factor is also observed with the increase in values of sterile angles. Hence more stringent bounds on sterile parameters are required to check the extent of CPT violation.
\FloatBarrier
\begin{figure}[htbp]
\begin{minipage}[b]{0.50\linewidth}
\centering
\includegraphics[width=\textwidth]{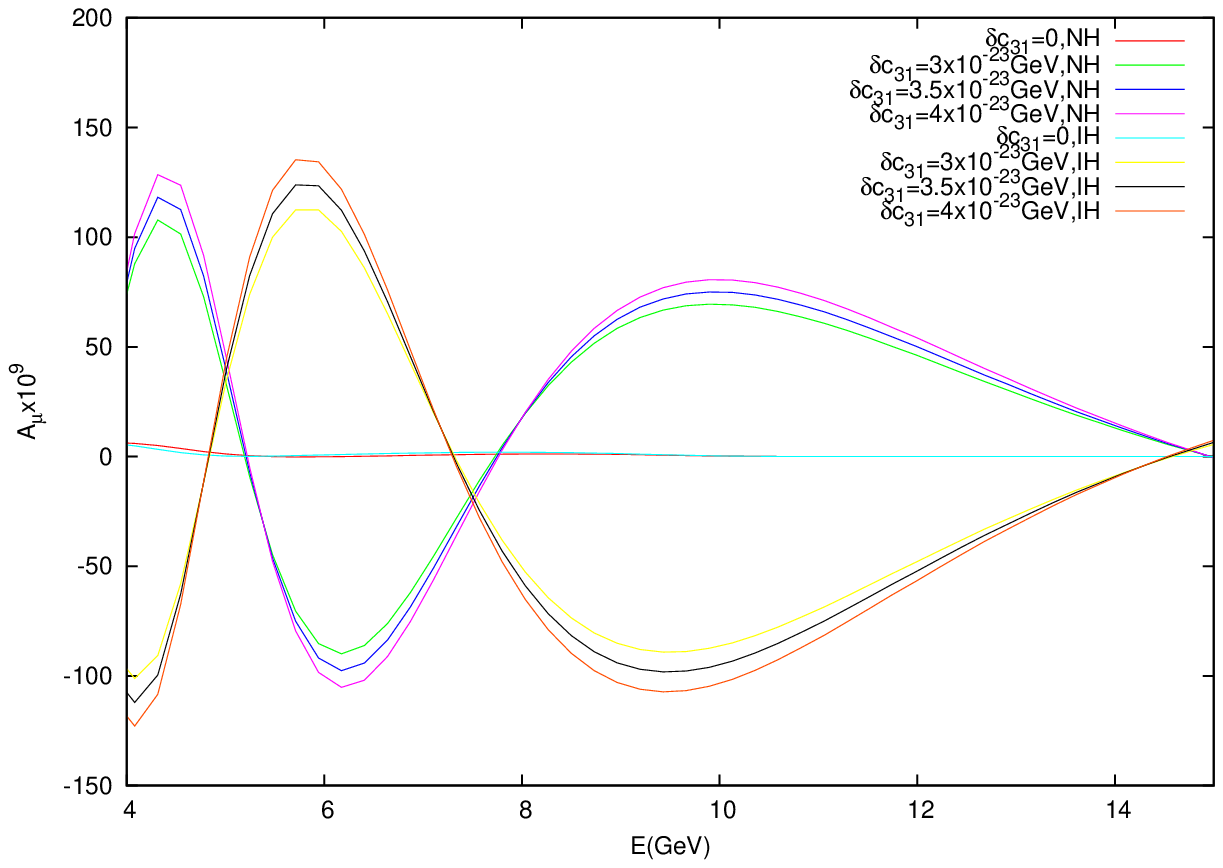}
\subfigure{(a)}
\end{minipage}
\hspace{0.7cm}
\begin{minipage}[b]{0.50\linewidth}
\centering
\includegraphics[width=\textwidth]{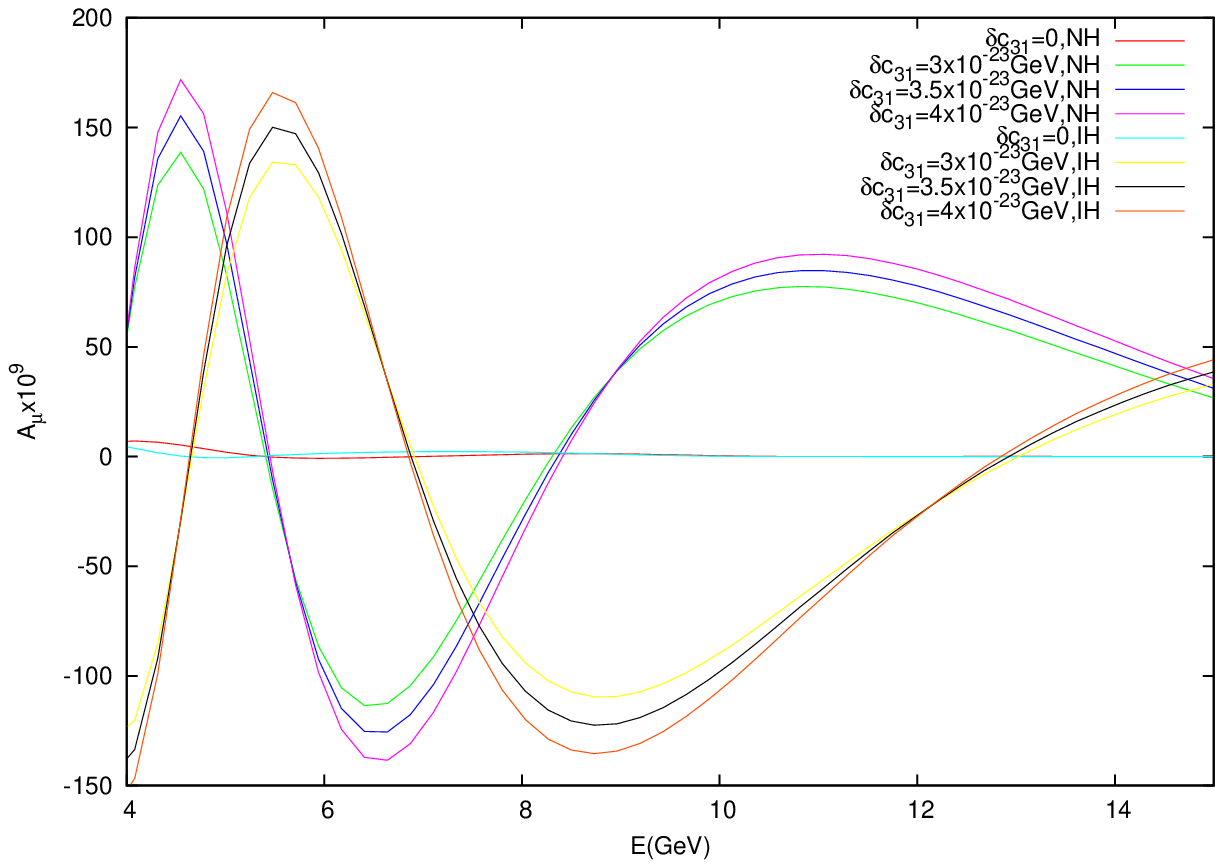}
\subfigure{(b) }
\end{minipage}
\caption{The variation in asymmetry factor $ A_{\mu} $(E) as a function of energy in energy range 4-15 GeV. Values of sterile parameters considered in  plot(a) and plot(b) are taken from long baseline experiments and reactor+atm experiments respectively
 }
\end{figure}
\FloatBarrier
\begin{figure}[htbp]
\begin{minipage}[b]{0.50\linewidth}
\centering
\includegraphics[width=\textwidth]{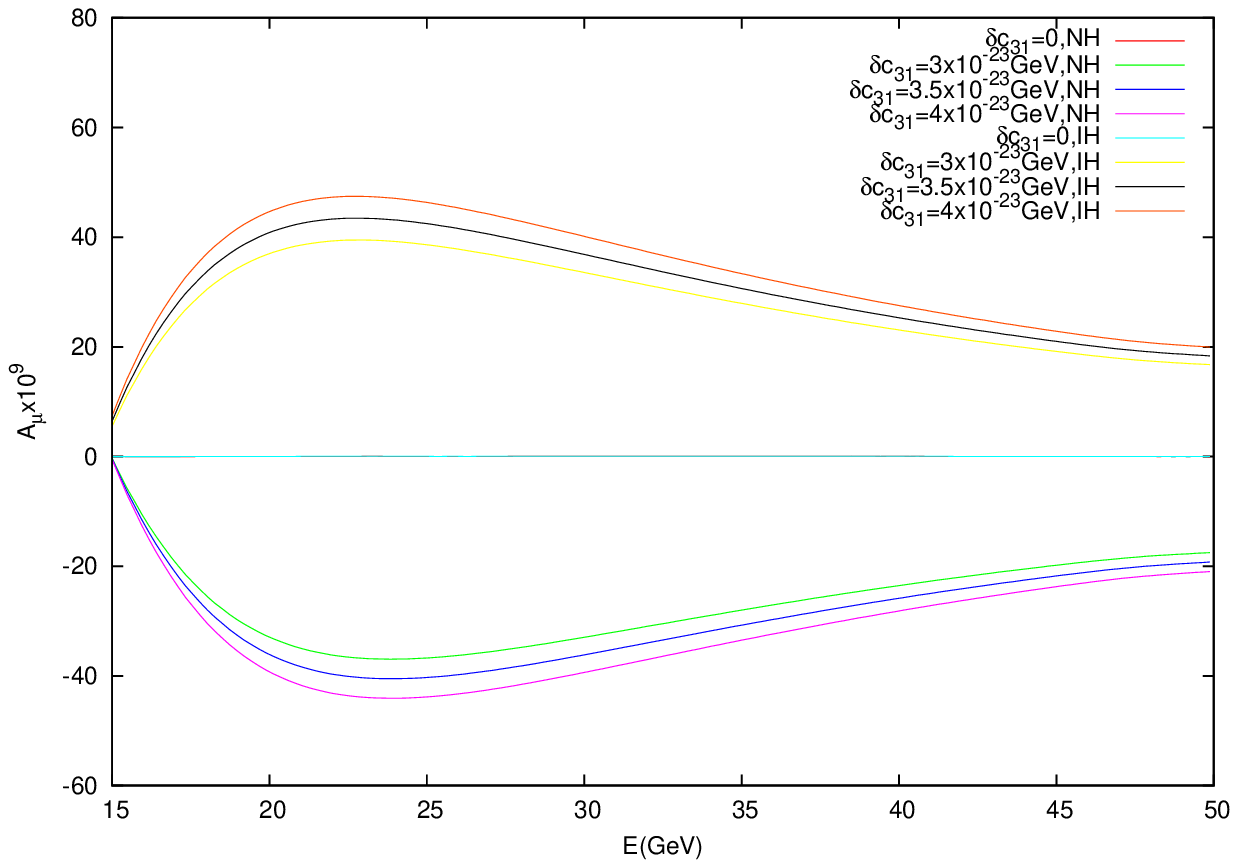}
\subfigure{(a)}
\end{minipage}
\hspace{0.7cm}
\begin{minipage}[b]{0.50\linewidth}
\centering
\includegraphics[width=\textwidth]{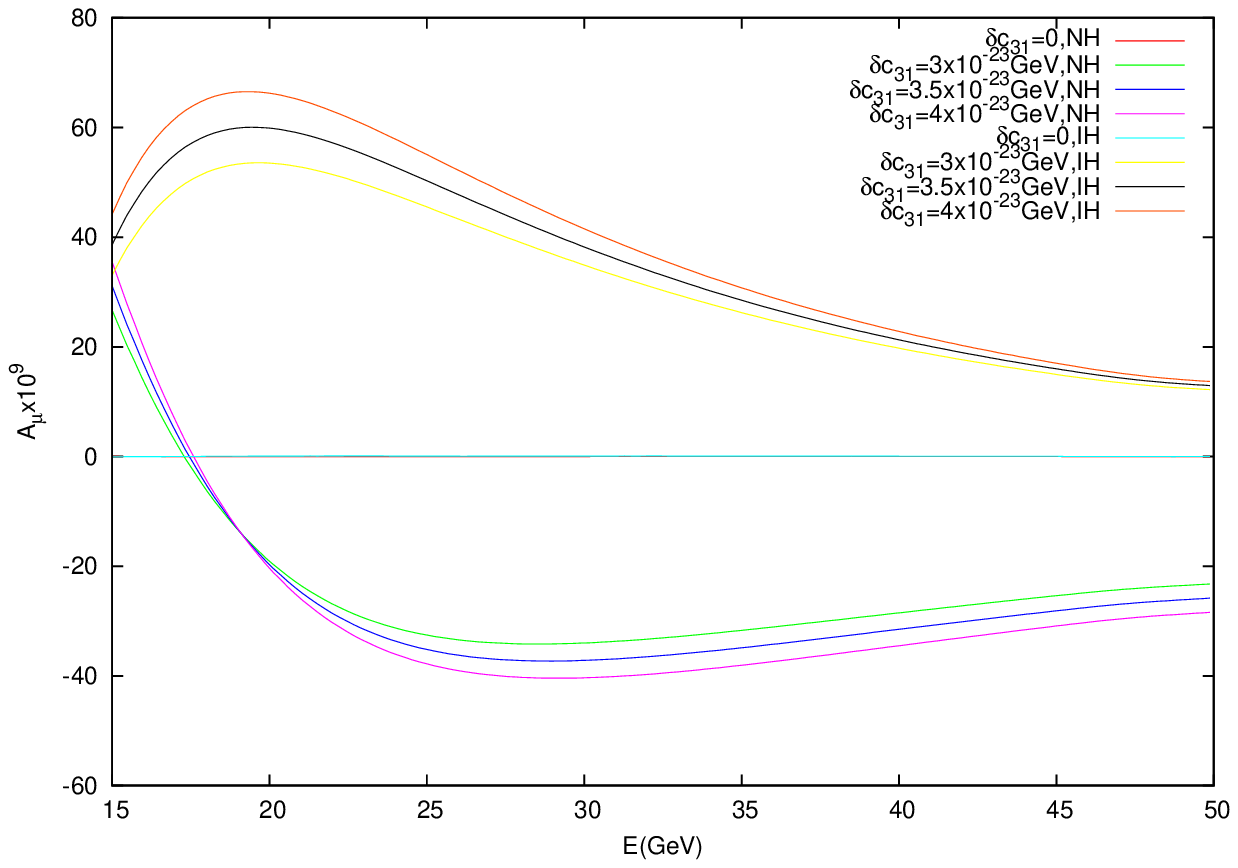}
\subfigure{(b)}
\end{minipage}
\caption{The variation in asymmetry factor $ A_{\mu} $(E) as a function of energy in energy range 15-50 GeV. Values of sterile parameters considered in plot(a) and plot(b) are taken from long baseline experiments and reactor+atm experiments respectively
 }
\end{figure}

\FloatBarrier
\begin{figure}[htbp]
\begin{minipage}[b]{\linewidth}
\centering
\includegraphics[width=0.70\textwidth]{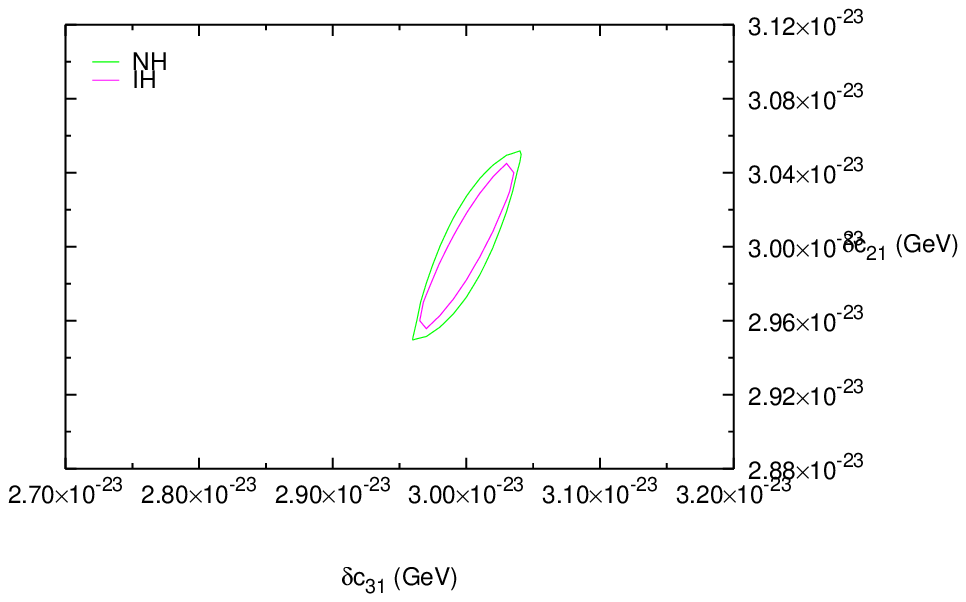}
\end{minipage}
\vspace{-1cm}
\begin{minipage}[b]{\linewidth}
\centering
\includegraphics[width=0.70\textwidth]{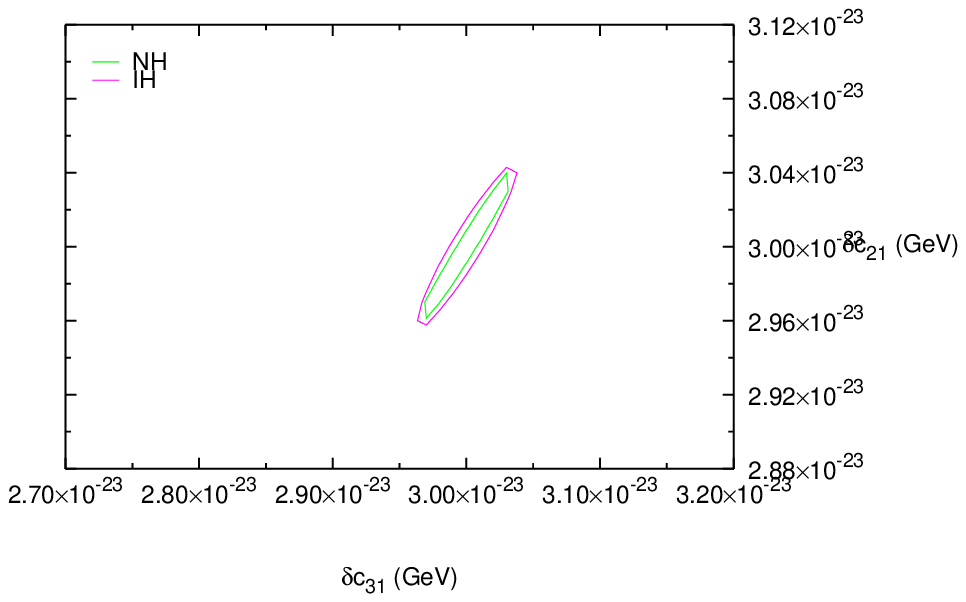}
\end{minipage}
\vspace{-1cm}
\begin{minipage}[b]{\linewidth}
\centering
\includegraphics[width=0.70\textwidth]{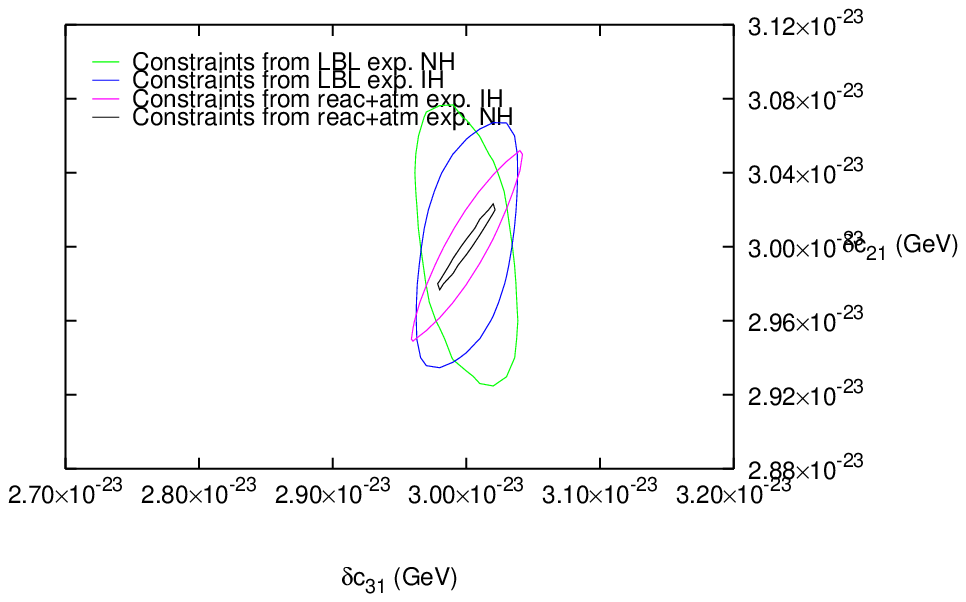}
\end{minipage}
\caption {The contours are plotted in $ \delta c_{31} $ -$ \delta c_{21} $ plane for 90\% C.L. with CPT violating terms $ \delta c_{31}$ = $ \delta c_{21} $ = $3.0\times10^{-23} $ GeV taking  energies 15 GeV,25 GeV and 50 GeV  respectively. In top two plots sterile parameters values are taken from  reactor+atmospheric experiments while for the bottom plot these values are taken from  both the reactor+atmospheric and long baseline experiments.}
\end{figure} 
\FloatBarrier
\begin{figure}[htbp]
\begin{minipage}[b]{\linewidth}
\centering
\includegraphics[width=0.70\textwidth]{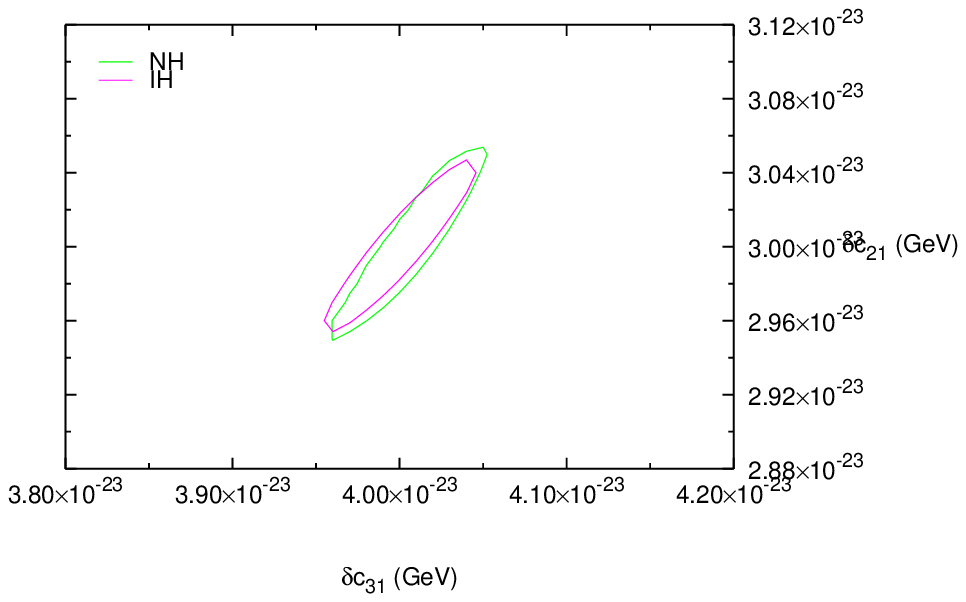}
\end{minipage}
\vspace{-1cm}
\begin{minipage}[b]{\linewidth}
\centering
\includegraphics[width=0.70\textwidth]{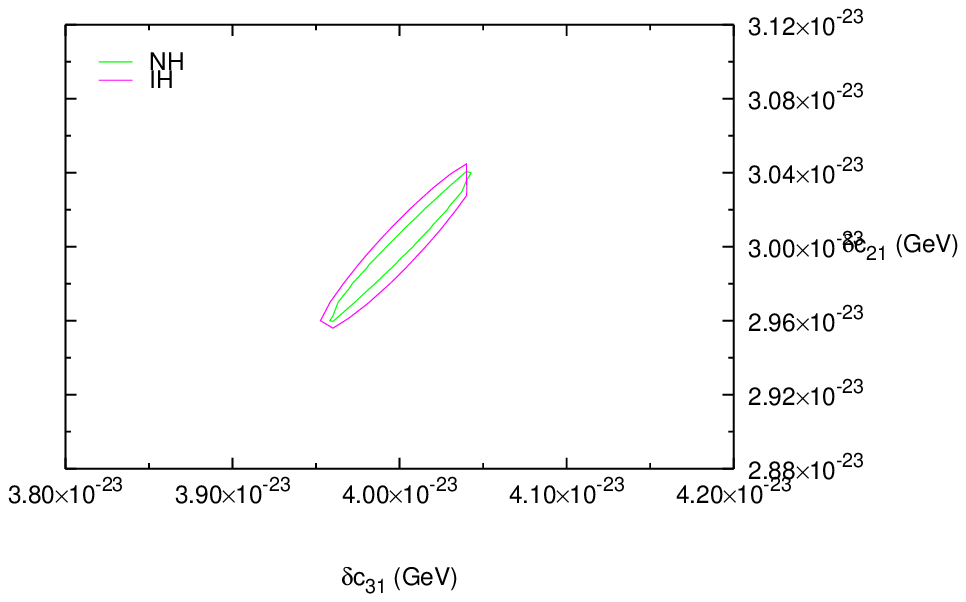}
\end{minipage}
\vspace{-1cm}
\begin{minipage}[b]{\linewidth}
\centering
\includegraphics[width=0.70\textwidth]{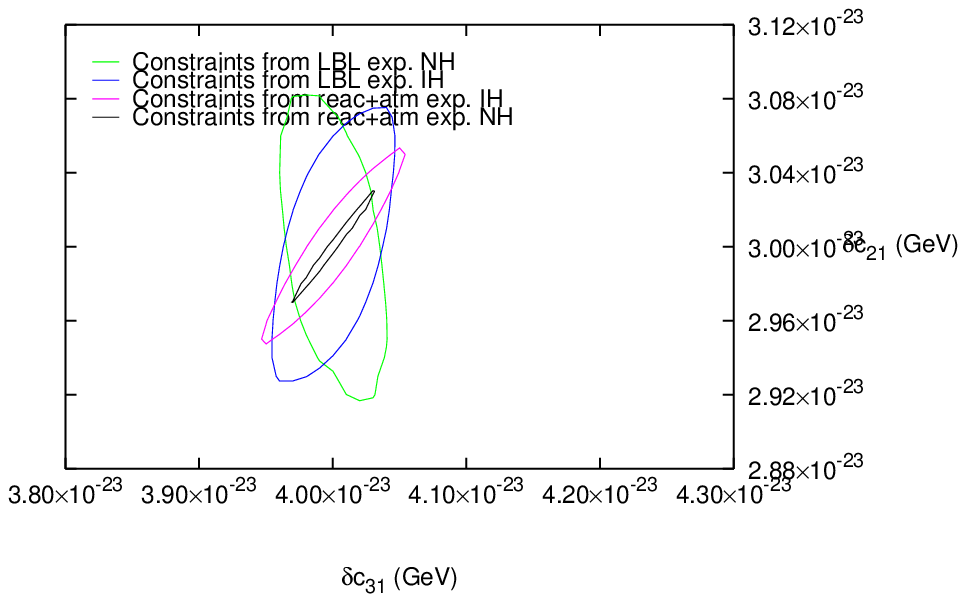}
\end{minipage}
\caption{The contours are plotted in $ \delta c_{31} $ -$ \delta c_{21} $ plane for 90\% C.L. with CPT violating terms $ \delta c_{31}$ = $4.0\times10^{-23} $ and $ \delta c_{21} $ = $3.0\times10^{-23} $ GeV taking  energies 15 GeV,25 GeV and 50 GeV  respectively. In top two plots sterile parameters values are taken from  reactor+atmospheric experiments while for the bottom plot these values are taken from both the reactor+atmospheric and long baseline experiments.}
\end{figure} 
\FloatBarrier
\begin{figure}[htbp]
\centering
\includegraphics[width=\textwidth]{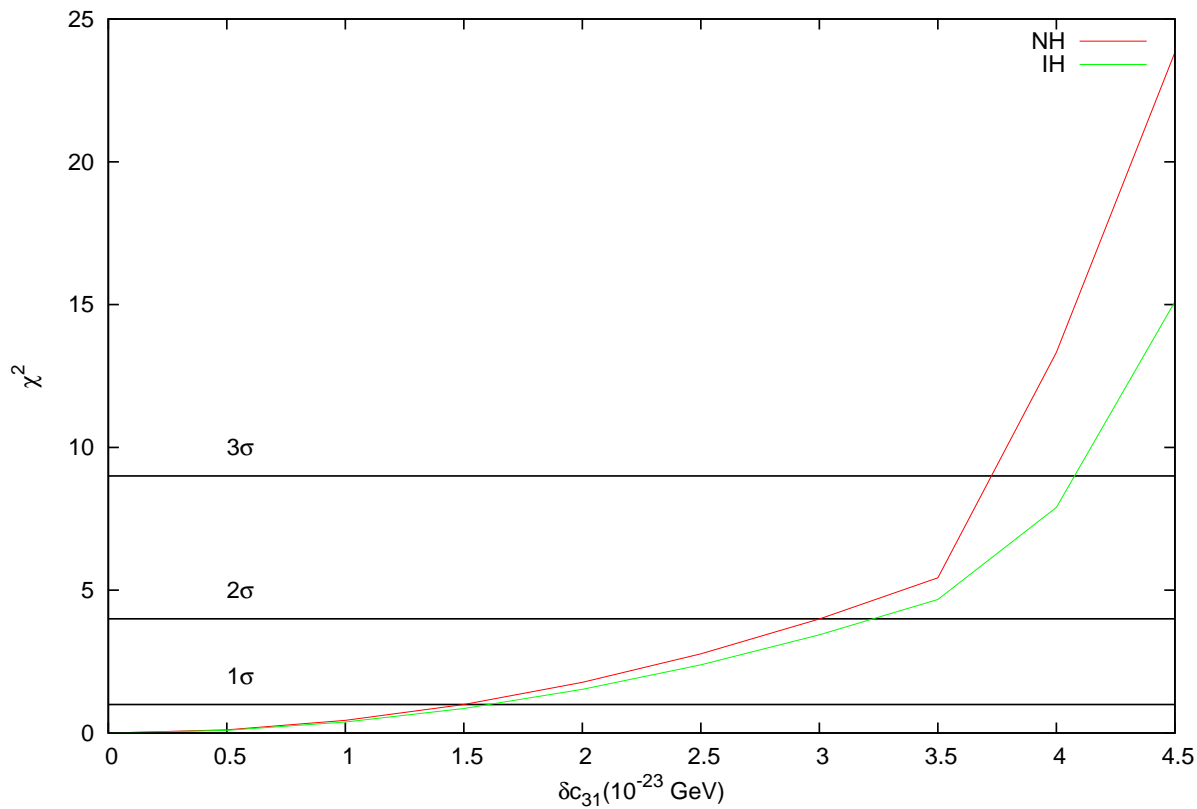}
\caption{$ \chi^{2}$ as a function of $ \delta c_{31} $ is shown. The green curve will represents inverted hierarchy and red curve represents normal hierarchy as true hierarchy. }
\end{figure}
In our work we have imposed bounds on CPT violating parameters $ \delta c_{31} $ and $ \delta c_{21} $ at 90\% C.L. Figures 10 and 11 demonstrate contours in  $ \delta c_{31} $ and $ \delta c_{21} $ plane with true value of CPT violating parameters as $ \delta c_{31} $ = $ 3\times 10^{-23} $ GeV,$ \delta c_{21} $ = $ 3\times 10^{-23} $ GeV and  $ \delta c_{31} $ = $ 4\times 10^{-23} $ GeV,$ \delta c_{21} $ = $ 3\times 10^{-23} $ GeV respectively. Each Figure consists of three plots at three different energies 15 GeV, 25 GeV and 50 GeV for baseline 7500 km. These plots illustrate bounds on CPT violating parameters at mentioned energies. The selection of three different energies are based on the results of previous observations (i.e R, $ \Delta R $ and $ A_{\mu} $). At selected energy 15 GeV, change in sign (+ve to -ve) is observed in the observables while studying effects of CPT violation for considered baselines and energies. This observation makes 15 GeV energy  important for studying CPT violating effects. A proposal of neutrino factory producing neutrino beam of 25 GeV muons is described in reference \cite{fermi} whereas in a different proposal we have a 50 GeV muon beam for the production of neutrinos in neutrino factory \cite{cern}. Therefore, by checking at the extent of bounds imposed on CPT violating parameters with energies 25 GeV and 50 GeV we want to check that by what order the results will improve if we move towards higher energies. By looking at different energy contours we conclude that amongst the three selected energies, 50 GeV energy is the best suited energy to constrain CPT violating parameter $ \delta c_{31} $, if nature allows NH to be true hierarchy. At the same time we observe that for long baseline experiment IH will be favourable hierarchy for determination of bounds on CPT violating parameters for energies less than 15 GeV. \\
As discussed earlier that, out of two parameters $ \delta c_{31} $ and $ \delta c_{21} $ considered in our analysis, the variation in $ \delta c_{31} $ will produce larger variation in the detectable  observables which are used in our work for checking CPT violation. Figure 12 shows value of $ \chi^{2} $ as a function of CPT violating parameter  $ \delta c_{31} $. It is plotted by marginalizing over oscillation parameters $ \Delta m_{31}^{2} $ in 3$ \sigma $ range of their best fit values  and $ \delta_{CP} $ from 0 to 2$ \pi $. Looking at figure we observe that the presence of CPT violation can be detected for  $\delta c_{31} \geqslant 3.6\times 10^{-23} $  GeV with neutrino factory for NH within 3$ \sigma $ limit.
 
 \section{Conclusions}
Neutrino factory will provide us a potential setup for observing T violation and setting significant bounds on CPT violation in neutrino sector. In four(3+1) neutrino flavor framework the angular mixing parameters of three active neutrinos are well constrained while the sterile parameters still needs better bounds on them. With the change in the value of sterile parameters a notable variation in bounds on CPT violating parameter and on the extent of T violation is captured by neutrino factory. Hence, well constrained values of sterile parameters will allow any  neutrino experiment to impose better constraints on T violation and CPT violating parameters. Amongst two selected sets of values of sterile parameters i.e. from long baseline experiments and reactor+atmospheric experiments we observed that neutrino factory potential for investigating T and CPT violation enhances when the sterile parameters values  are equal to those which are constrained by reactor and atmospheric experiments. Neutrino factory with 50 GeV energy is sensitive to probe T violation when true values of sterile parameters  will be equal to those predicted by reactor+atmospheric experiments. We stipulate that a pure CPT violating effects can be observed along short baseline i.e 1300 km-2000 km with energies 4 GeV to 6 GeV where extrinsic CPT violation  is negligible. On the other hand at long baselines we can observe these effects with energies in the range 20 GeV- 40 GeV along baselines 4000 km-7500 km. CPT violating parameters $\delta c_{31} \geqslant 3.6\times 10^{-23} $ GeV for NH and $\delta c_{31} \geqslant 4\times 10^{-23} $ GeV for IH will make neutrino factory capable to capture signatures of CPT violation at 3 $ \sigma $ level.

\begin{appendices}
\numberwithin{equation}{section}
\section{ Eigenvalues and Eigenvectors of Hamiltonian to second order }
 Using the time independent perturbation theory we calculate the eigenvalues and eigenvectors of hamiltonian $H_{f}$ up to the order of $\eta^{2}$ correctly.\\ 
Eigenvalues  of $H_{0}$ are given by  \vspace{-0.5cm}
\begin{center}
\begin{equation}
 E_{1}^{(0)}  =a_{e}+a_{n}, E_{2}^{0}=a_{n} , E_{3}^{(0)}  =a_{e}+1 , E_{4}^{(0)}  =\sigma \end{equation}
\end{center}
where $ \sigma = \dfrac{\bigtriangleup m^{2}_{41}}{\bigtriangleup m^{2}_{31}}$ and $a_{e,n}\equiv \dfrac{A_{e,n}}{\bigtriangleup m^{2}_{31}}  $\\ \\
Eigenvectors of $ H_{0} $ are given by
\begin{center}
\begin{equation}
 V_{1}^{(0)}=\begin{bmatrix}
1\\
0\\
0\\
0\\
\end{bmatrix}  ,  V_{2}^{(0)}=\begin{bmatrix}
0\\
-c_{23}\\
s_{23}\\
0\\
\end{bmatrix}  ,  V_{3}^{(0)}=\begin{bmatrix}
0\\
s_{23}\\
c_{23}\\
0\\
\end{bmatrix}  ,  V_{4}^{(0)}=\begin{bmatrix}
0\\
0\\
0\\
1\\
\end{bmatrix} 
\end{equation}
\end{center}
Eigenvalues and eigenvectors  for $ H_{1} $ are calculated by using equations (A.3) and (A.4) respectively.
\vspace{-6ex}   
\begin{center}
\begin{equation}
 E_{j}^{(1)}= < V_{j}^{0}\mid H_{1}\mid V_{j}^{0}> 
\end{equation}
\end{center}
\vspace{-6ex} 
\begin{center}
\begin{equation}
 \mid V_{j}^{(1)}>=   \displaystyle\sum_{k\neq j}\mid V_{k}^{0}>\dfrac{ < V_{j}^{0}\mid H_{1}\mid V_{j}^{0}>}{E_{j}^{(0)}-E_{k}^{(0)}}
\end{equation}
\end{center}
Eigenvalues and eigenvectors of $H_{2}$ can be calculated with the help of zeroth and first order eigenvalues and eigenvectors mentioned in equations (A.5) and (A.6).  given as \vspace{-5ex}
 \begin{center}
\begin{equation}
 E_{j}^{(2)}= < V_{j}^{0}\mid H_{1}\mid V_{j}^{1}> 
\end{equation}
\end{center}
\vspace{-5ex} 
 \begin{center}
\begin{equation}
 \mid V_{j}^{(2)}> = \dfrac{-\mid V_{j}^{(0)} >}{2} \displaystyle\sum_{k\neq j}\dfrac{\mid V_{jk} \mid^{2}}{E^{2}_{kj}}+\displaystyle\sum_{k\neq j}\mid V_{k}^{(0)}>\left[ \displaystyle\sum_{k^{\shortmid} \neq j} \dfrac{V_{k j^{\shortmid} } V_{k^{\shortmid} j}}{E_{kj}E_{k^{\shortmid} j}} -\dfrac{V_{kj}V_{jj}}{E_{kj}^{2}}\right] 
\end{equation}
\end{center}
where 

$V_{kj}=<k^{0} \mid V \mid j^{0} > $ and $ E_{kj}=E_{k}^{0}-E_{j}^{0} $\\
The total eigenvalues and eigenvectors of $ H_{f} $ up to second order is given by \vspace{-5ex} 
\begin{center}
\begin{equation}
E_{total}=E_{j}^{0}+E_{j}^{1}+E_{j}^{2} 
\end{equation}
\end{center}
\vspace{-6ex} 
\begin{center}
\begin{equation}
V_{total}=V_{j}^{0}+V_{j}^{1}+V_{j}^{2}
\end{equation}
\end{center}
Using the set of four normalized eigenvectors we form the unitary matrix $\tilde{U}$ as.\vspace{-4ex}

\begin{center}
\begin{equation}
\tilde{U} = \begin{bmatrix}
(V_{1m})_{1} & (V_{1m})_{2} & (V_{1m})_{3} & (V_{1m})_{4}\\
(V_{2m})_{1} & (V_{2m})_{2} & (V_{2m})_{3} & (V_{2m})_{4}\\
(V_{3m})_{1} & (V_{3m})_{2} & (V_{3m})_{3} & (V_{3m})_{4}\\
(V_{4m})_{1} & (V_{4m})_{2} & (V_{4m})_{3} & (V_{4m})_{4}\\

\end{bmatrix}
\end{equation}
\end{center}
where $ V_{jm}$ is  normalized vector.\\
Now hamiltonian $ H_{f} $ can be diagonalised by using  the above derived unitary matrix $ \tilde{U} $ and the diagonalized hamiltonian $ H_{D} $ can be expressed as  \vspace{-8ex}
\begin{center}
\begin{equation}
 H_{D} = \tilde{U}^{\dagger} H_{f}\tilde{U} 
\end{equation}
\end{center}
\end{appendices} 
 \begin{abstract}
One of the authors Sujata Diwakar is thankful to University Grant Commission,India  for giving financial support under the Rajiv Gandhi National Fellowship scheme.
\end{abstract}

\end{document}